\documentclass{article}
\usepackage{booktabs} % For formal tables
\usepackage[margin=1in]{geometry}
\usepackage{amsmath,amssymb, graphicx,multicol,multirow,array, bbm,varwidth,semantic}
\usepackage{amsbsy,amsfonts,amsthm,commath,graphicx}
\usepackage[english]{babel}
\usepackage{microtype}      % microtypography
\usepackage{graphicx}
\usepackage[utf8]{inputenc} % allow utf-8 input
\usepackage[T1]{fontenc}    % use 8-bit T1 fonts
\usepackage{microtype}      % microtypography
\usepackage{mathtools}

\usepackage{hyperref}       % hyperlinks
\usepackage{url}            % simple URL typesetting
\usepackage{booktabs}       % professional-quality tables
\usepackage{nicefrac}       % compact symbols for 1/2, etc.
\usepackage{microtype}      % microtypography
\usepackage{graphicx,multicol,multirow,bbm,array,varwidth,semantic} %amssymb 
\usepackage{mathtools}

\usepackage[english]{babel}
\usepackage{verbatim}

\usepackage{siunitx}

\usepackage{xcolor}
\usepackage{enumitem}

\def\eqref#1{equation~(\ref{#1})}
\def\1{\bm{1}}

\DeclareMathAlphabet{\mathsfit}{\encodingdefault}{\sfdefault}{m}{sl}
\SetMathAlphabet{\mathsfit}{bold}{\encodingdefault}{\sfdefault}{bx}{n}

\newcommand{\E}{\mathbb{E}}

\DeclareMathOperator{\Ind}{\mathbbm{1}}

\usepackage{tikz}

\usetikzlibrary{shapes,decorations,arrows,calc,arrows.meta,fit,positioning}
\tikzset{
    -Latex,auto,node distance =1 cm and 1 cm,semithick,
    state/.style ={ellipse, draw, minimum width = 0.7 cm},
    point/.style = {circle, draw, inner sep=0.04cm,fill,node contents={}},
    bidirected/.style={Latex-Latex,dashed},
    el/.style = {inner sep=2pt, align=left, sloped}
}

\usepackage{color}

\definecolor{darkgreen}{rgb}{0.0, 0.5, 0.0}

\usepackage{natbib}
\usepackage{soul}

\title{Incentives and Outcomes in Bug Bounties}
\date{}
\author{Serena Wang\footnotemark[1]~\footnotemark[2] \quad \quad Martino Banchio\footnotemark[1]~\footnotemark[3] \quad \quad Krzysztof Kotowicz\footnotemark[1] \\  Katrina Ligett\footnotemark[1]~\footnotemark[4] 
\quad \quad R. Preston McAfee\footnotemark[1]
\quad \quad Eduardo’ Vela” Nava\footnotemark[1] \\
	\\
 	\small{\footnotemark[1]~Google Research} \\
	\small{\footnotemark[2]~Department of Computer Science, University of British Columbia} \\
    \small{\footnotemark[3]~Department of Economics and Innocenzo Gasparini Institute for Economic Research, Università Bocconi} \\
	\small{\footnotemark[4]~School of Computer Science and Engineering and Federmann Center for the Study of Rationality,}\\
	\small{Hebrew University} 
}
\begin{document}

% Title page for title and abstract only.
% \begin{titlepage}

\maketitle

% Optionally include a table of contents
% \vspace{1cm}
% \setcounter{tocdepth}{1} % adjust to 1 if desired
% \tableofcontents

% \end{titlepage}

% Abstract. Note that this must come before \maketitle.
\begin{abstract}
Bug bounty programs have contributed significantly to security in technology firms in the last decade, but little is known about the role of reward incentives in producing useful outcomes. We analyze incentives and outcomes in Google’s Vulnerability Rewards Program (VRP), one of the world’s largest bug bounty programs. We analyze the responsiveness of the quality and quantity of bugs received to changes in payments, focusing on a change in Google’s reward amounts posted in July, 2024, in which reward amounts increased by up to 200\% for the highest impact tier. Our empirical results show an increase in the volume of high-value bugs received after the reward increase, as well as a high positive observed elasticity of labor supply for such bugs. We further break down the sources of this increase between veteran researchers and new researchers, showing that the reward increase both redirected the attention of veteran researchers and attracted new top security researchers into the program.

\end{abstract}

% Paper body

\section{Introduction}

Bug bounty programs, which reward external researchers for reporting security vulnerabilities, have become a ubiquitous part of the cybersecurity landscape.  Both for tech giants running their own programs and for smaller firms outsourcing to specialized platforms, bug bounty programs are seen as a crucial tool for enhancing security \citep{ellis2022bounty}. Following recent growth in large-scale AI systems, bug bounty programs have seen even further renewed interest as a tool for supporting AI safety, as companies like Google, Microsoft, Anthropic, and Open AI have launched new programs in the last several years \citep{openai_bug_bounty, google_ai_bug_bounty}.
% Bug bounty programs have seen even further recent growth: Recent concerns about safety in large-scale AI systems have led to a renewed interest in utilizing third-party information elicitation to support security.
% They are part of a growing trend towards utilizing third-party expertise for information elicitation, a practice that also extends to areas such as AI development.  

The rise of bug bounty programs is not an isolated phenomenon, but rather part of a broader transformation of the nature of work and expertise.  These programs mirror the growing trend towards leveraging the ``gig economy,'' where individuals contribute their skills and knowledge on a project-by-project basis, rather than through traditional employment \citep{mas2019labor,jeronimo2021gig}.  Bug bounty programs tap into a global pool of independent, highly competent researchers, creating a flexible and scalable approach to security and an intriguing complement to in-house security teams. The decentralized, incentive-driven nature of bug bounty programs, therefore, offers a valuable lens through which to examine the impact of the freelance work model on specialized fields. Moreover, the use of external ``red teams'' and bug bounty programs in areas like AI safety further underscores the shift away from traditional employment models towards leveraging external expertise for critical tasks \citep{feffer2024red}. As these crowdsourcing and gig-based mechanisms continue to evolve, continual empirical research into their effectiveness---particularly concerning incentive structures and the quality of output---becomes increasingly vital.

Despite the widespread adoption of bug bounty programs, fundamental questions remain about how these programs actually work, particularly regarding the role of financial incentives.  Do larger rewards lead to the discovery of more critical vulnerabilities? How do incentives affect different types of researchers? These questions apply not just to the design of software bug bounty programs, but also to the design of information elicitation mechanisms across domains, including AI evaluation and online crowdsourcing. 

In this paper, we empirically study these questions in the context of Google's Vulnerability Rewards Program (VRP), one of the world's leading bug bounty initiatives. We leverage a particularly clean dataset collected at an opportune time in 2024 when a significant increase in reward amounts was deployed by Google (up to a 200\% increase), but few exogenous changes were happening (e.g. external price changes or the use of AI to find bugs). We statistically treat this as an event study, and the timing helps us provide a clean snapshot of the bug bounty market, which is useful both in its own right and as a baseline for future research as the landscape continues to change.

% We estimate the responsiveness of the quality and quantity of bugs received to changes in the incentive structure by exploiting a major increase in Google's reward levels, of up to 200\% for the highest impact tier.
% Our results show that the volume of high-value reports increased significantly in response to the reward increase. We then study the channels responsible for this increase.
% We compute a significant, positive elasticity of labor supply. We show that these changes are generated by veteran researchers and new researchers alike: the reward increase both redirected the attention of existing participants towards higher-value targets, and attracted new top talent into the program.
% We show that much of these changes are generated by the top security researchers, who tend to file more, higher-impact reports. Finally, the extensive margin seems to dominate the intensive margin, suggestive of a competitive bug bounty marketplace.

Our analysis focuses on two questions that are core to the operation and outcomes of a bug bounty program. First, we ask, \textit{How do changes to the reward incentives affect the quantity and quality of bugs received?}
We observe a shift in distribution towards high quality bugs. We also compute a significant, positive elasticity of labor supply, particularly in high-value bugs.
This suggests room for growth in high-value bugs and possible saturation in low-hanging fruit. 
A second important question regards participation -- \textit{how do security researchers (existing and new) respond to reward changes}? We show that these changes are generated by veteran researchers and new researchers alike: the reward increase both redirected the attention of existing participants towards higher-value targets, and attracted new top talent into the program. 
A subtlety is that the reward change did not increase the total number of new researchers, but instead attracted a small number of highly productive researchers, suggesting substitution in a competitive labor market. 
Overall, our statistical analysis of outcome data yields insights into the effects of increasing rewards on security outcomes, researcher productivity, and attraction of new researchers, all of which are key to informing operational decisions, and build a foundation for future theoretical and empirical research.
%We show that top security researchers responded more strongly to increased reward incentives than less productive, incidental participants. 

% Section \ref{sec:background} gives background on bug bounties and a detailed description of the operation of Google's Vulnerability Rewards Programs. We describe the reward increase that occurred in July, 2024, and give important details on the assembly of the dataset. Section \ref{sec:methods} provides a conceptual framework and details the statistical methods used to test for changes in quantity. Presentation of results begins in Section \ref{sec:all_bugs} with all bugs, followed by a breakdown in Section \ref{sec:high-value} by bug type and in Section \ref{sec:top_researchers} by researcher type. We synthesize our results in Section \ref{sec:conclusions}, including discussion of policy implications, limitations, and future work.
\subsection{Related Work.}
% There is limited prior work studying economic incentives in bug bounty programs.
% While prior work has studied bug bounty programs from a qualitative, theoretical, and empirical standpoint, there is still limited work demonstrating the effects of economic reward incentives in a large centralized program such as the Google VRP.
%Our work leverages a unique significant policy change that occurred in July of 2024 in which Google tripled its rewards. This adds new perspective by both studying a previously understudied type of program, using a specific recent event with few other external events that might confound the analysis. The data is also particularly expressive, as we have data on actual bugs found and their severity, rather than just report rates. Thus, we're able to compute elasticities and study security researcher retention where previous studies have not.
% Our work focuses on economic incentives, and adds new perspective through statistical analysis of responses to a change in rewards in a large bug bounty program, focusing on actual bugs found instead of reports.\snote{TODO: add new references, add more on difference b/w google program and others, qualitative additional contributions from theoretical angle}
% - Type of bug bounty program: Our setup adds value in studying an in-house program run by a large company, focused on 
% - Uniqueness of data:
From an empirical standpoint, 
%our work joins a line of prior studies of various data sources and programs \citep{alexopoulos2021vulnerability,ruohonen2018bug}.  
several works \citep{finifter2013empirical,zhao2015empirical,luna2019productivity,sridhar2021hacking,piao2024study,alexopoulos2021vulnerability,ruohonen2018bug} have measured researcher productivity and vulnerability discovery rates across various data sources and programs (e.g. Google, Firefox, HackerOne, and BugCrowd). Limitations from observational data have led to a limited ability to make causal interpretations, as well as an inability to compute fine-grained elasticities. 
%A notable exception is the work of \citet{sridhar2021hacking}, which leverages proprietary data from the HackerOne platform to compute a price elasticity of supply. However, also due to data limitations, \citet{sridhar2021hacking} were unable to account for the severity of vulnerabilities in their analysis. 
Our study has the unique advantage of leveraging a significant event in which Google tripled its rewards, allowing us to estimate outcome changes and elasticities across vulnerability types in a quasi-experimental setup. To our knowledge, our study is the first to estimate a price elasticity of supply for vulnerabilities stratified by impact and severity. 
%providing a quasi-experimental setup in which we 
%It is also important to note that third-party bug-bounty platforms like that studied by \citet{sridhar2021hacking} differ qualitatively in many ways from large in-house programs like the Google VRP, and previous work \citep{luna2019productivity} has suggested that insights may not generalize between the two.

% Previously, \citet{sridhar2021hacking} computed elasticities across the HackerOne platform, and take an instrumental variables approach to causality. However, \citet{sridhar2021hacking} were unable to incorporate the severity of vulnerabilities into their analysis. Thus, we are the first to report elasticities stratified by vulnerability severity.

% similar outcomes in Google Chrome and Firefox bug bounty programs in 2013. In addition to analyzing more recent large-scale programs, our study has the added advantage of focusing on a significant increase in rewards at a single point in time in 2024, allowing us to measure changes and estimate causal effects. \citet{luna2019productivity} analyze quantities of bugs reported in HackerOne data and in publicly released Google VRP data; however, they do not include any analysis of payments or responsiveness to incentives. The HackerOne program also differs significantly from Google's, in that the platform serves many smaller companies, and the types of bugs paid for tend to be less severe. 
% \citet{piao2024study} empirically study cooperation behavior among security researchers, which is an interesting but separate question from those we address. 

From a theoretical standpoint, \citet{gal2024merchants} model incentives and effort for different types of hackers, comparing expert white hat, non-expert white hat, and black hat hackers. Our work address related questions empirically through an event study, and also finds that different types of security researchers emerge naturally in the data. 

Qualitative interview and survey-based methods have previously been used to study bug hunters' motivations \citep{akgul2023bug,fulton2023vulnerability,telang2025balancing} and bug bounty programs as a whole \citep{alomar2020you,ellis2022bounty,laszka2018rules}. Most relatedly, \citet{akgul2023bug} survey bug hunters to gain insight into the factors that they find most motivating and challenging. They identify multiple important benefits and programmatic challenges, among which rewards incentives are one of the most important motivating factors. Our work quantifies the realized impacts of changes in reward incentives on outcomes.
% Our work focuses on reward incentives, and complements such qualitative analysis by working directly with outcome data. 
% Among aspects of bug bounty programs outside of reward incentives, \citet{telang2025balancing} have criticized lack of transparency of some bug bounty programs, recommending standards for reporting and disclosure to avoid delays in patching vulnerabilities. We focus on the relationship between rewards and the vulnerability discovery process.

% Beyond bug bounty programs, \citet{dellago2022exploit} characterize the landscape of exploit brokers, which may sell bugs to actors seeking to exploit bugs rather than fix them. We focus on bug bounty programs in which the purpose is to improve product security.

% \textcolor{red}{Beyond bug bounty programs, incentives for crowdwork have been empirically studied more generally \cite{ho2015incentivizing,yin2014monetary}. For instance, Ho et al. \cite{ho2015incentivizing} ran an experiment varying bonus payments for crowdworkers on Amazon Mechanical Turk, finding that increasing rewards is effective if workers believe that a task is ``effort-responsive.'' Our study provides additional evidence in a live bug bounty program, where we find that an increase in rewards led to increased findings of high-value bugs.}

We contribute to the literature on alternative work arrangements (see \citet{mas2020alternative} for a review) by studying a global labor marketplace with a number of distinguishing characteristics. 
First, the work schedule is completely flexible, and no explicit or implicit contract is in place for the researcher's time. The most interesting aspect is perhaps the stochastic and highly heterogeneous mapping between a worker's effort and that worker's compensation, generated by heterogeneity in worker skill and background; an element of luck; and the uncertainty regarding a bug's existence and the time necessary to discover it. We join other empirical studies on incentives for crowdwork \citep{ho2015incentivizing,yin2014monetary} including that of \citet{ho2015incentivizing}, who varied bonus payments for crowdworkers on Amazon Mechanical Turk and found that increasing rewards is effective if workers believe that a task is ``effort-responsive.'' Our study provides additional evidence in a live bug bounty program, where we find that an increase in rewards led to increased findings of high-value bugs.

Our work also contributes to the literature  estimating elasticities of labor supply. Many quasi-experimental studies exploit changes in tax rates to estimate elasticities of labor supplies, using macro data (such as the work of \citet{prescott2004americans} and subsequent work). Instead we leverage microeconomic data on a change in the reward structure in a fully-flexible labor market. One subtlety of the analysis lies in labor substitution patterns between different bug bounty programs: among large bug bounty programs, only Google's rewards increased in the period of interest, implying changes in the relative pay rates across different programs.

% TODO: cut to 1 page
\section{Setting and Background on Bug Bounties}\label{sec:background}

% \paragraph{Definition and overview of existing bug bounty programs}
Our study focuses on bug bounty programs, which we define as programs hosted by a firm (e.g., a technology company) that offer rewards to external participants for discovering bugs in the technology systems of the firm for the purposes of improving security. Such programs are  hosted by many large technology companies, including Google, Meta, Amazon, Microsoft, Mozilla, and Apple. We focus here on bug bounty programs with the purpose of improving security, hosted by the firm itself, rather than exploit brokers or third-party platforms.
% In addition to large firms running their own bug bounty programs, there also exist third-party platforms such as HackerOne and BugCrowd that run bug bounty programs on behalf of client companies. Another, related, marketplace is that of various \textit{exploit brokers}, which offer rewards for bugs and then resell them, potentially to actors who seek to exploit them. We focus here on bug bounty programs with the purpose of improving security, hosted by the firm itself.
% We focus on bug bounty programs with the purpose of improving security, rather than \textit{exploit brokers}, which offer rewards for bugs that they then sell to potentially malicious actors. 

\subsection{Background on Google's Vulnerability Rewards Programs (VRPs)} To gain insight into the incentives and outcomes of such bug bounty programs, we analyze proprietary data from Google's Vulnerability Rewards Programs (VRPs). We give an overview of how Google's bug bounty programs operate, and document our methods for assembling the dataset in detail to support reproducibility in this and other bug bounty programs.

Google has multiple separate vulnerability rewards programs, roughly corresponding to product areas. For example, there are separate programs for Chrome, Android, Abuse, Open Source Software, etc. Each program has its own reward policies, with can vary in reward amounts and in the types of bugs prioritized. Our primary subject of study is the \textit{Google and Alphabet Vulnerability Rewards Program (GAVRP)}, which deployed a major reward increase in July 2024.
% In general, the GAVRP serves core products not covered by the other more targeted programs. 
We describe the reward increase in more detail in Section \ref{sec:reward_increase}. As a point of comparison, we also analyze data from several Google VRPs whose rewards have not changed in the last several years.
% : the \textit{Open Source Software Vulnerability Rewards Program (OSSVRP)} and the \textit{Abuse Vulnerability Rewards Program (AVRP)}.

%Figure \ref{fig:pas} shows the distribution of product areas for bugs found through the GAVRP. 
% In October 2024, the \textit{Cloud Vulnerability Rewards Program (CVRP)} was launched as a separate entity after having previously been combined with the \textit{Google and Alphabet Vulnerability Rewards Program}. The CVRP has had its own separate rewards table and review panel, but provides similar reward amounts to the new reward policy of the GAVRP after July, 2024. In the main paper, we treat the CVRP and GAVRP data as a single program, as they have continuously addressed similar bugs as a whole with similar reward amounts at each point in time. We discuss this more detail in Section \ref{sec:cloud}.

\textbf{Rewards.} Each VRP publishes a \textit{rewards table} as well as rules by which bugs are evaluated.\footnote{GAVRP rewards table: \url{https://bughunters.google.com/about/rules/google-friends/6625378258649088/google-and-alphabet-vulnerability-reward-program-vrp-rules}} This table lists reward amounts for bugs of specified tiers (roughly, the severity of the potential impact of a vulnerability in this domain) and categories (roughly, the type of vulnerability, e.g., remote code execution, unrestricted database access, etc.). The rewards tables list the maximum rewards possible for different bug types; however, final reward determinations are made by reviewers within Google, and may involve bonuses or penalties as listed in each VRP's rules.\footnote{\url{https://bughunters.google.com/about/rules/}}  Examples of considerations that result in bonuses and penalties include attack limitations, exploitablity of the reported vulnerability, and the quality of the submitted bug report. 

\textbf{Participants.} We refer to participants who submit to the bug bounty programs as \textit{security researchers}, or \textit{researchers} for short. These participants may be employed at other companies, may look for bugs full-time, or are sometimes in academia. The Google VRPs have international participation -- in 2023, a total of 632 security researchers from 68 countries were paid across all programs \citep{vrp_blog_2023}.

\textbf{Life cycle of a report.} The entry point for a security researcher to participate in the VRP is to submit a vulnerability \textit{report}. All submitted reports go through an internal triage process in which Google-internal security engineers evaluate the report on metrics such as report quality, vulnerability severity, and vulnerability impact. 
Figure \ref{fig:report_status} shows various evaluation stages for a report.
We associate each bug with the date its corresponding report was submitted.

A bug report can spawn multiple \textit{product bugs}, as Google-internal reviewers funnel issues found in a bug report to relevant product teams. Most reports do not generate any product bugs. Some product bugs are identified as duplicates of each other if they are found to be linked to the same underlying problem. As this work focuses on the \textit{outcomes} of bug bounty programs, we analyze a de-duplicated set of product bugs that are an end result of this triaging process; for simplicity, we refer to these resulting items simply as \emph{bugs}. 
% Analysis of raw report data may also be of separate interest in future work.
We restrict our analysis to product bugs specifically classified as \textit{vulnerabilities} by program reviewers, as these are a primary target of the Google VRPs. 
% Bugs of other types may sometimes be identified from a report, and may be classified as feature requests or other general bugs that do not pose a direct security issue.

\subsection{Reward Increase and Dataset Summary}\label{sec:reward_increase}

In July 2024, the GAVRP deployed a significant reward increase which increased the rewards posted for top tier bugs by approximately 200\%. Before this, rewards in this program had not substantially changed since 2013. 
% \footnote{We include screenshots of posted rewards tables from before and after the reward increase in an extended version of the paper at \url{https://arxiv.org/abs/2509.16655}.}
Figure \ref{fig:reward_tables} shows posted rewards tables from before and after the reward increase in July 2024.
We focus on analyzing immediate changes in outcome in the six months following the reward increase, using a dataset that consists of all de-duplicated distinct product bugs classified as vulnerabilities received between January 2023 and December 2024. This time period has the advantage of avoiding confounding effects from some external factors over time, such as price changes from other companies \citep{openaiincrease,anthropic,microsoftzeroday,appleincrease}, and a later rise in popularity of AI tools for bug reports in 2025 \citep{aireports}. We discuss extended effects into September, 2025 in the Appendix.

% We leverage data from before and after this reward increase to study the effects of reward incentives on program outcomes and researcher productivity. Below we describe in detail the studied programs and our methods for processing the accompanying data.

\textbf{Treated program.}\label{sec:cloud} We define the \textit{treated program} to be the combined submissions made to the GAVRP and the Cloud Vulnerability Rewards Program (CVRP), which branched from the GAVRP in October, 2024. The combination of these two programs covered the same software targets in the periods before and after the reward increase, and were both subject to the same reward increase.\footnote{We note that there may have been an effect from the \textit{announcement} of the new Cloud program when it branched off from the GAVRP. As a robustness check, we repeat our analyses with all Cloud-related bugs removed in Appendix \ref{app:nocloud} (and find similar outcomes).}
Our dataset for the \textit{treated program} consists of a total of 957 bugs from 487 distinct researchers.

% after having previously been combined with the GAVRP \citep{cloud_intro}. The CVRP has its own separate rewards table (Figure \ref{fig:reward_table_cloud}), which provides similar reward amounts to the post-July-2024 reward policy of the GAVRP. %The CRVP covers bugs that would have previously been under the GAVRP. 
% Our primary analysis combines bugs from the CRVP and GAVRP together and continues to treat these as a single combined program, as the combined programs have continuously covered the same bug submissions with similar reward amounts at all points in time. 
% We refer to the combination of the GAVRP and CVRP as the \textit{treated program}.%, as these programs together offered similar rewards for a similar collective class of bugs throughout the period of study. 

% The main limitation of this approach is that it is difficult to disentangle the effect of the \textit{announcement} of the new Cloud program. After all, any growth in submissions observed in or soon after October 2024 could also be due to a delayed effect of higher effort invested after the reward increase in July, as it is common for new bugs to take months to find. As a robustness check, we repeat our analyses with all Cloud-related bugs removed in Appendix \ref{app:nocloud} (and find similar outcomes).
%, and perform additional experiments showing that there is not currently a statistically significant effect of the announcement of Cloud-related bugs in October.

\textbf{Untreated programs} We also leverage data from Google VRPs that did not change their rewards during the period of study: the Abuse Vulnerability Rewards Program (AVRP) and the Open Source Software Vulnerability Rewards Program (OSSVRP), both of which have had similar reward amounts since 2022. We refer to both of these as the \textit{untreated programs.} Our datasets for the two \textit{untreated programs} consists of a combined total of 367 bugs from 199 distinct researchers. There is very little overlap of researchers between the treated and untreated programs, and the effects of between-program transfer are negligible.\footnote{We provide between-program transfer statistics in Appendix \ref{app:substitution}.}

\section{Conceptual Model and Statistical Methodology}\label{sec:methods}

We introduce the core conceptual model and the statistical methods employed to analyze changes in the quantity of bugs received and to compute elasticities with respect to reward amounts. We later extend this notation to address questions specific to bug types. 

% empirically analyze three main elements: changes in quantity of bugs received, changes in distribution of bugs types received, and elasticities with respect to reward amounts. 

% \subsection{Tests for changes in quantity}

% Our first group of statistical tests addresses questions of whether the change in rewards in July 2024 was associated with changes in the quantity of bugs received.

% We first analyze changes in quantity of bugs received. The outcome of interest is the rate of submission of bugs, measured as the number of bugs submitted per month.

% Our first group of statistical tests addresses questions of whether the change in rewards in July 2024 was associated with changes in the quantity of bugs received. 
For each month $t \in \mathcal{T}$, we observe a bug count $y^t_{g}$ for program $g$. Let $t^{*} \in \mathcal{T}$ denote the month in which the reward change was deployed, i.e., $t^{*} = $ July 2024. Let $Y^{0}_{g}, Y^{1}_{g}$ be random variables representing the observed rate of bugs received per month for program group $g$ under the previous reward policy and under the increased reward policy, respectively. 
% Throughout the main paper we consider only three programs, $g \in \{\text{treated}, \text{untreated 1}, \text{untreated 2}\}$.
%, where $g =  $ represents the untreated programs (bugs from AVRP and OSSVRP), and $g = \text{treated}$ represents the treated programs (bugs from GAVRP and CVRP).

\textbf{Basic change in mean.}
We obtain empirical estimates for the mean rate of bugs found per month before and after the reward change as 
\[\bar{Y}^{0}_{g} =\frac{\sum_{t \in \mathcal{T}}y^t_{g} \Ind(t < t^{*})}{\sum_{t \in \mathcal{T}} \Ind(t < t^{*})}, \quad \bar{Y}^{1}_{g} =\frac{\sum_{t \in \mathcal{T}}y^t_{g} \Ind(t \geq t^{*})}{\sum_{t \in \mathcal{T}} \Ind(t \geq t^{*})}.\]

An unpaired two-sided t-test tests the null hypothesis that $\E[\Delta Y_{g}] = 0$, where $\Delta Y_{g} = Y^{1}_{g} - Y^{0}_{g}$.

\textbf{Regression discontinuity designs.}
% A causal estimand of interest is the average treatment effect at the time of the reward change, which can be found using regression discontinuity designs. 
% This is a distinct causal estimand from that of the DiD setup, but still illustrative.
To estimate a causal effect of the reward increase at the time of deployment, we apply a sharp \textit{regression discontinuity design (RDD)} over data from the treated program. The units of observation are defined to be outcomes observed in a given month. Let $Y$ denote a random variable representing the number of bugs per month, and let $T$ denote the month, where $T=0$ for the month of the initial reward change deployment (July, 2024), and otherwise takes the value of the number of months before or after July 2024 (so $T = 1$ for August 2024, and $T = -1$ for June 2024, etc.). A unit is treated for all months in our data including and after July 2024: $D \coloneq \Ind(T \geq 0)$. Applying the Neyman-Rubin potential outcomes framework \citep{neyman1923applications,rubin1974estimating}, let $Y(0), Y(1)$ represent the potential outcome for the number of bugs that would be received in a given month under the previous reward policy and under the new reward policy, respectively. The goal is to estimate the local average treatment effect: $\tau^{\text{RDD}} = \E[Y(1) - Y(0) | T = 0]$.
The core identification assumption is local continuity: $\lim_{t \uparrow 0} \E[Y|T = t] = \E[Y(0)|T = t]$, and $\lim_{t \downarrow 0} \E[Y|T = t] = \E[Y(1)|T = t]$.
% \begin{align*}
%     \lim_{t \uparrow 0} \E[Y|T = t] = \E[Y(0)|T = t], \quad \lim_{t \downarrow 0} \E[Y|T = t] = \E[Y(1)|T = t].
% \end{align*}
% \[\lim_{t \uparrow 0} \E[Y|T = t] = \E[Y(0)|T = t], \]
% \[\lim_{t \downarrow 0} \E[Y|T = t] = \E[Y(1)|T = t].\]
Thus, our estimator is $\hat{\tau}^{\text{RDD}} = \hat{\mu}^{1} - \hat{\mu}^0$,
where $\hat{\mu}^{1}$ estimates $\lim_{t \downarrow 0} \E[Y|T = t]$, and $\hat{\mu}^{0}$ estimates $\lim_{t \uparrow 0} \E[Y|T = t]$.

In practice, there may be a delayed effect of the reward change on the observed outcomes due to the fact that bugs can take months to find, even after a researcher starts putting effort into looking. We also consider a \textit{regression kink design (RKD)}, where the goal is to estimate the local change in slope for the outcome. This would better capture an increase in effort near the time of the reward increase.
Our estimand of interest under the regression kink design is the change in \textit{slope} of the bug count per month before and after the reward change: $\tau^{\text{RKD}} = \lim_{t \downarrow 0} \frac{d\E[Y(1) | T = t]}{dt} - \lim_{t \uparrow 0} \frac{d\E[Y(0) | T = t]}{dt}$.
 % \[\tau^{\text{RKD}} = \lim_{t \downarrow 0} \frac{d\E[Y(1) | T = t]}{dt} - \lim_{t \uparrow 0} \frac{d\E[Y(0) | T = t]}{dt}.\]

We estimate both $\hat{\tau}^{\text{RKD}}$ and $\hat{\tau}^{\text{RDD}}$ via a local linear regression of the form
\[Y = \alpha + \beta_{0}T + \beta_{1}D + \beta_{2} DT + \varepsilon, \]
\noindent over bandwidth $T \in C$. An estimate for $\hat{\tau}^{\text{RDD}}$ is given by the regression estimate for the coefficient $\beta_{1}$, and an estimate for $\hat{\tau}^{\text{RKD}}$ is given by the regression estimate for the coefficient $\beta_{2}$. We provide 
% both conventional ordinary least squares (OLS) estimates for $\beta_{1}$ and $\beta_{2}$, 
 bias-corrected estimates with robust confidence intervals for $\beta_{1}$ and $\beta_{2}$ following the methodology of  \citet{calonico2014robust}. We select the MSE-optimal bandwidth according to \citet{calonico2020optimal}. 

\textbf{Chow test.} 
Drawing from timeseries analysis, we perform a Chow test to test for a change point in the rate of bugs per month at the fixed time point of July 2024. A Chow test is a hypothesis test which, in principle, indicates that individual regressions before and after the change are a better fit than a single regression for the full time period.
We assume linear models before and after the change, where the rate of bugs per month is $Y = \alpha_{0} + \beta_{0}T + \varepsilon$ for $T < 0$, and $Y = \alpha_{1} + \beta_{1}T + \varepsilon$ for $T \geq 0$.
% \[Y = \alpha_{0} + \beta_{0}T + \varepsilon \; \text{ for } \; T < 0, \quad Y = \alpha_{1} + \beta_{1}T + \varepsilon \;\text{ for } \; T \geq 0.\]
The null hypothesis is that $ \alpha_{0} =  \alpha_{1}$ and $ \beta_{0} =  \beta_{1}$. An F-test yields the test statistic.

\textbf{Elasticities.} 
We next turn to estimating the \textit{elasticity} of labor supply, defined as the rate of change in the quantity of bugs found in comparison to the rate of change in the rewards. This gives measures of how \emph{responsive} bug reporting is to changes in incentives.

Let $\eta$ denote a point elasticity, defined as the ratio of the percent change in quantity to the percent change in reward: $\eta = \frac{\%\Delta Y}{\%\Delta R}$, where $\%\Delta Y = \frac{E[Y^1 - Y^0]}{E[Y^0]}$ and $\%\Delta R = \frac{E[R^1 - R^0]}{E[R^0]}$. 
% \[\eta = \frac{\%\Delta Y}{\%\Delta R}, \; \text{ where } \; \%\Delta Y = \frac{E[Y^1 - Y^0]}{E[Y^0]}, \;  \%\Delta R = \frac{E[R^1 - R^0]}{E[R^0]}.\]
$Y^0, Y^1$ are random variables denoting bugs received per month in the time period before and after the reward change, respectively, and $R^0, R^1$ are random variables denoting reward given per bug in the time period before and after the reward change, respectively. We denote sample mean counterparts with a bar (e.g. $\bar{Y}^0$ indicates the sample mean counterpart of $Y^0$), and $\hat{\eta}$ as the elasticity computed from the sample means: $\hat{\eta} = \frac{\%\Delta \bar{Y}}{\%\Delta \bar{R}}$, where $\%\Delta \bar{Y} = \frac{\bar{Y}^1 - \bar{Y}^0}{\bar{Y}^0}$ and $\%\Delta \bar{R} = \frac{\bar{R}^1 - \bar{R}^0}{\bar{R}^0}$. 
% \[\hat{\eta} = \frac{\%\Delta \bar{Y}}{\%\Delta \bar{R}} \;  \text{ where } \; \%\Delta \bar{Y} = \frac{\bar{Y}^1 - \bar{Y}^0}{\bar{Y}^0}, \;  \%\Delta \bar{R} = \frac{\bar{R}^1 - \bar{R}^0}{\bar{R}^0}.\]
Note that we measure $\bar{R}^0, \bar{R}^1$ as the average \textit{realized} rewards paid per bug, and not the maximum possible reward published in the reward tables. To understand the variation in our elasticity estimate $\hat{\eta}$, we compute a two-sided bootstrap 95\% confidence interval based on 500 Monte Carlo resamples with replacement from the dataset of bugs.

\textit{Remark.} Causal interpretation of our elasticity estimates can be confounded by any external changes outside of the Google VRP rewards that would affect researcher effort in the program, including, e.g., price changes in external bug bounty program, or advances in bug-finding tooling. To our knowledge in the study period, we did not find significant changes to other major external competing programs. We also restricted the end of the study period to 2024, to mitigate confounding from the adoption of increasingly powerful AI tools in 2025. Still, we do not exclude the possibility of confounding.
\section{Observed effects on quantity and quality of bugs received}\label{sec:all_bugs}

\begin{figure*}[!ht]
    \centering
    \begin{tabular}{cc}
    \begin{minipage}{0.47\textwidth}
    \centering
    \begin{tikzpicture}
      \node (img)  {\includegraphics[scale=0.4]{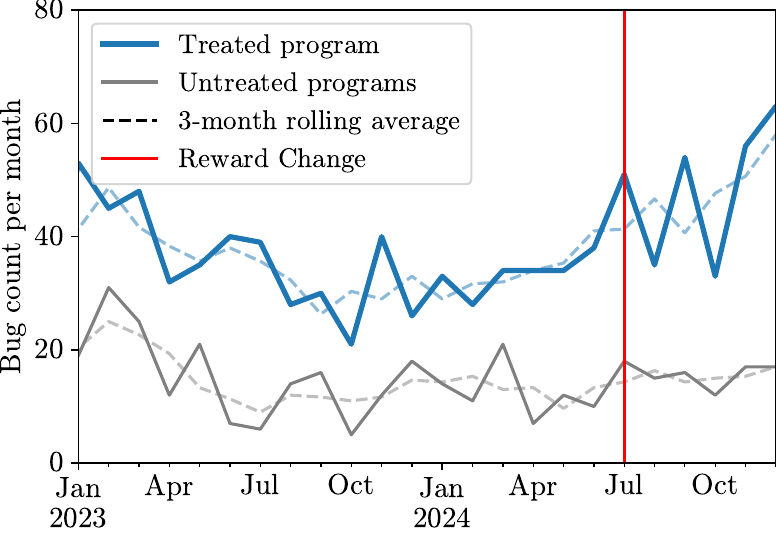}};
      \node[above=of img, node distance=0cm, rotate=0, anchor=center,yshift=-1cm,xshift=0.2cm] {\footnotesize{Bug counts in treated vs. untreated programs}};
     \end{tikzpicture}
    \end{minipage} & \begin{minipage}{0.47\textwidth}
    \centering
    \begin{tabular}{c|c|c|c}
    \toprule
        Statistic & Value & 95\% CI & p-value \\
        \midrule
        $\Delta \bar{Y}$ & \textbf{12.94} & \textbf{(3.86, 22.02)} & \textbf{0.007} \\
        % $\hat{\tau}^{\text{DiD}}$ & \textbf{12.39} & \textbf{(1.85, 22.93)} & \textbf{0.022} \\
        $\hat{\tau}^{\text{RDD}}$ & 10.08 & (-8.01, 28.16) &  0.27  \\
        $\hat{\tau}^{\text{RKD}}$ & -11.04 & (-30.07, 7.99) & 0.25 \\
        Chow test & N/A & N/A & \textbf{0.004
        } \\
    \bottomrule
    \end{tabular}
    \end{minipage}
    \end{tabular}
    \caption{
    % Visualizations of bug counts and regressions for all bugs in the treated and untreated programs. 
    \textit{(Left)} All bugs per month received in the treated and untreated programs. \textit{(Right)} Quantity change statistics for all bugs from the treated program. Rows with p-value $< 0.05$ are in bold.}
    \label{fig:all_timeseries}
\end{figure*}

We begin with a birds-eye view of the changes in overall bug counts before and after the reward increase in July, 2024. Figure \ref{fig:all_timeseries} shows all bugs per month received, from the treated and untreated programs, as well as statistics for the change in quantity for all bugs from the treated program. The observed change in bugs received per month, $\Delta \bar{Y}$, indicates that on average, $12.94$ more bugs per month were received after the reward change, which is roughly a 37\% increase from before.

Overall, total bug counts increased on average, but there is not definitive statistical evidence of a causal effect on total bug count. However, this does not mean that the reward increase did not have \textit{any} causal effects; in fact, as will soon observe, outcomes varied significantly by bug type, with much stronger effects for high-value bugs.

% Applying the statistical methods outlined above, we report results for changes in quantity and estimates of elasticity for all bugs in the treated vs.~untreated programs. 
% We will break this down into bugs of different types in Section \ref{sec:high-value}.

% \begingroup
% \setlength{\tabcolsep}{-2pt} 
% \endgroup

% \begin{figure*}[!ht]
%     \centering
%     \includegraphics[scale=0.4]{figures/all_timeseries_combined.pdf}
%     \caption{All bugs per month received from 2023-2024.}
%     \label{fig:all_timeseries}
% \end{figure*}

% \begin{figure}[!ht]
%     \begin{minipage}{0.5\textwidth}
%     \centering
%     \begin{tikzpicture}
%       \node (img)  {\includegraphics[scale=0.5]{figures/all_timeseries_gc.pdf}};
%       \node[above=of img, node distance=0cm, rotate=0, anchor=center,yshift=-0.9cm,xshift=0cm] {\footnotesize{Treated program}};
%      \end{tikzpicture}
%     \end{minipage}
%     \caption{All bugs per month received from 2023-2024 in the treated program.}
%     \label{fig:all_timeseries}
% \end{figure}
% \begin{figure}[!ht]
%     \begin{minipage}{0.5\textwidth}
%     \centering
%     \begin{tikzpicture}
%       \node (img)  {\includegraphics[scale=0.5]{figures/all_timeseries_ao.pdf}};
%       \node[above=of img, node distance=0cm, rotate=0, anchor=center,yshift=-0.9cm,xshift=0cm] {\footnotesize{Untreated program}};
%      \end{tikzpicture}
%     \end{minipage}
%     \caption{All bugs per month received from 2023-2024 in the untreated program.}
%     \label{fig:untreated_timeseries}
% \end{figure}

\subsection{Different effects for high-value bugs}\label{sec:high-value}

% A primary security goal of a bug bounty program is to produce is to reveal ``high-value'' bugs to the firm, where ``high-value'' could be defined in many ways, from high exploitability, to the involvement of highly sensitive attack targets, to bugs that would have otherwise been difficult for internal engineers to find. Thus, when analyzing the impact of incentives, a practical question of interest is: \textit{How do changes to incentives affect the distribution of high-value bugs received?} 

% The reward change targeted specific types of bugs, with up to a 3$\times$ increase for Tier 0 bugs (from $\$ 31,337$ to $\$101,010$), and no change for Tier 3 bugs. A natural question is then, \textit{Are more high-tier bugs received after the reward table change?}

% More generally, the primary security goal is to increase the rate of submission of high-value, ``important'' bugs. Thus, the overall goal of this section is to ask, \textit{Did the reward change effectively increase receipt of important bugs?}

% Further, \textit{How much does increased investment improve the quality of bugs received?}

The primary security impact concern is whether the reward change increased the receipt of \textit{high-value} bugs, rather than simply increasing the volume of less impactful, ``low hanging fruit'' type bugs. 
%More specifically, a primary security goal of a bug bounty program is to reveal ``high-value'' bugs to the firm, 
``High-value'' could be defined in many ways, from high exploitability, to the involvement of highly sensitive attack targets, to bugs that would have otherwise been difficult for internal engineers to find. 
%Thus, we analyze quantity changes both in all bugs and for high-value bugs types.
The reward increase in the treated program seems to reflect an intention to incentivize the submission of high-value bugs, as maximum rewards increased roughly 200\% for Tier 0 bugs (from a maximum reward of $\$ 31,337$ to $\$101,010$), while there was no change in rewards for Tier 3 bugs. 
%Thus, we may also expect that the effects of such a change would be different per tier. 

Thus in this section, we analyze the impacts of the reward increase on different bug types, with a particular focus on high-value bugs.
We consider several proxies available in the data for ``high-value'' bugs, namely \textit{tier}, \textit{severity}, and \textit{merit}. Specifically,
% , denoted by the random variables $X_{\text{T}}, X_{\text{S}}, X_{\text{M}}$, respectively. Specifically,
\begin{itemize}[topsep=1pt,itemsep=0pt]
    \item The \textit{tier} represents the domain in which the vulnerability is found, where Tier 0 represents domains with global impact. 
    % Tiers make up the columns in the rewards table.
    % \footnote{https://bughunters.google.com/about/rules/google-friends/6625378258649088/google-and-alphabet-vulnerability-reward-program-vrp-rules}
    Categories are $\mathcal{X}_{\text{T}} = $ \{Tier 0, Tier 1, Tier 2, Tier 3, None\}. `None' indicates that no tier was assigned to the bug, which often corresponds with no reward. 
    \item The \textit{severity} variable is given by an annotation created by the VRP reviewers, with categories $\mathcal{X}_{\text{S}} = \{\text{Critical, High, Medium, Low, None}\}$. Tier and severity are possibly correlated but not qualitatively the same---tier indicates the attack target, and severity is a property of the attack itself. A value of `None' indicates that no severity score was assigned to the bug.
    \item The high \textit{merit} indicator is a combination of annotations created by the VRP reviewers that indicate reports of exceptional quality. Categories $\mathcal{X}_{\text{M}} = \{\text{High, None}\}$ indicate whether a bug received a high merit annotation or not.
\end{itemize}

We provide descriptive statistics and hypothesis tests for distributional differences, as well as tests for causality that rely on the specified identification assumptions. In addition to considering the quantity and distribution of bugs submitted, we also account for the actual rewards given, and compute estimated price elasticities. We use a subscript $x$ on estimands, estimators, and random variables as shorthand to indicate conditioning on bugs belonging to category $x$.
%-- e.g., $\tau^{\text{RDD}}_{x_{\text{T}}} = \E[Y(1) - Y(0) | T=0, X_{\text{T}} = x_{\text{T}}] $ indicates the treatment effect for bugs with tier $x_{\text{T}}$.

\textbf{Observed changes in distribution.}\label{sec:distribution}
% We run a basic hypothesis test for a change in the distribution of types for bugs received before and after the reward change. 
We first measure observed changes in the distribution over bug types of the bugs received by the treated program. We begin with a basic hypothesis test for whether the distribution has shifted over type categories for each type variable. For each type variable, let $N_x^0, N_x^1$ denote the number of bugs of category $x$ observed before and after the reward change. We conduct a $\chi^2$ test of independence where the null hypothesis is that the counts $\{N_x^0: x \in \mathcal{X}\}$, $\{N_x^1: x \in \mathcal{X}\}$, are drawn from the same multinomial distribution over categories $\mathcal{X}$. Figure \ref{fig:type_hists} shows results of this test alongside normalized histograms of the tiers for all bugs received through the treated program before and after the reward change in July 2024. The results suggest distribution shift for both Tier and Merit. 
\footnote{We give an expanded multiple hypothesis testing setup in Appendix \ref{app:multiple_hypothesis_test} to test for shifts towards and away from each category, which demonstrates significant shifts towards Tier 0 and High Merit, as well as a significant shift away from Tier 2.}

\begingroup
\begin{figure*}[!h]
\setlength{\tabcolsep}{0pt}
    \centering
    \begin{tabular}{ccp{0.25\linewidth}}
    % \centering\small{Tier} & \begin{center}\small{Severity} \end{center} & \small{Merit}  \\
    \begin{minipage}{0.38\textwidth}
    \centering
    \begin{tikzpicture}
      \node (img)  {\includegraphics[scale=0.35]{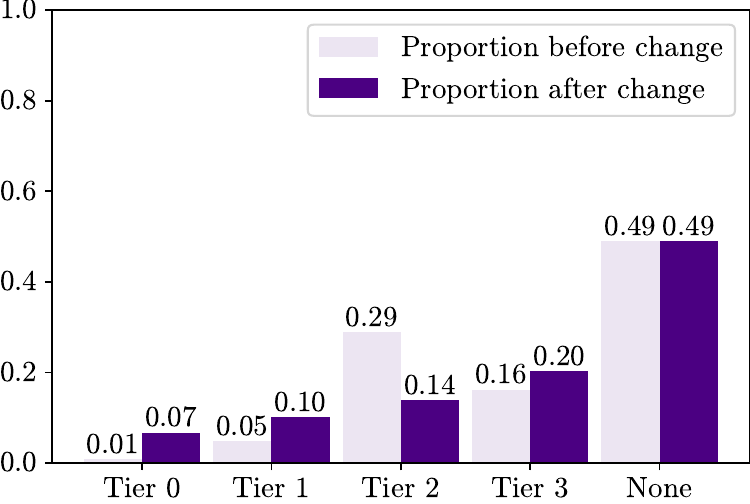}};
      \node[above=of img, node distance=0cm, rotate=0, anchor=center,yshift=-0.8cm,xshift=0.2cm] {\small{Tier}};
     \end{tikzpicture}
    \end{minipage} & 
    \begin{minipage}{0.38\textwidth}
    \centering
    \begin{tikzpicture}
      \node (img)  {\includegraphics[scale=0.35]{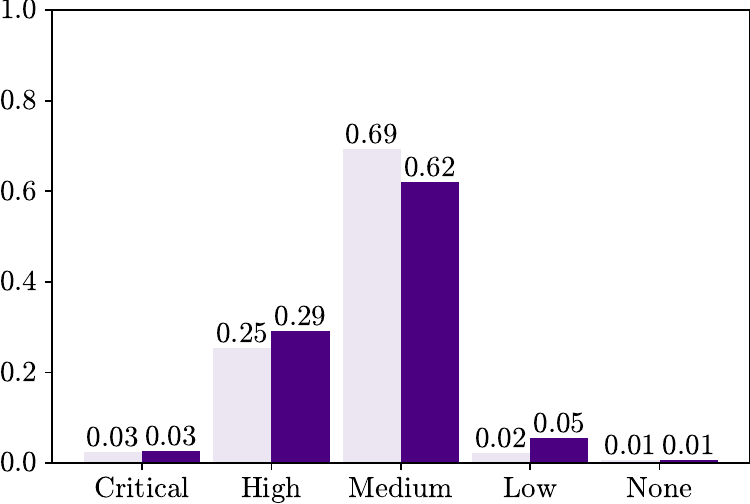}};
      \node[above=of img, node distance=0cm, rotate=0, anchor=center,yshift=-0.8cm,xshift=0.2cm] {\small{Severity}};
    \end{tikzpicture}
    \end{minipage} &
    \begin{minipage}{0.25\textwidth}
    \centering
    \begin{tikzpicture}
      \node (img)  {\includegraphics[scale=0.28]{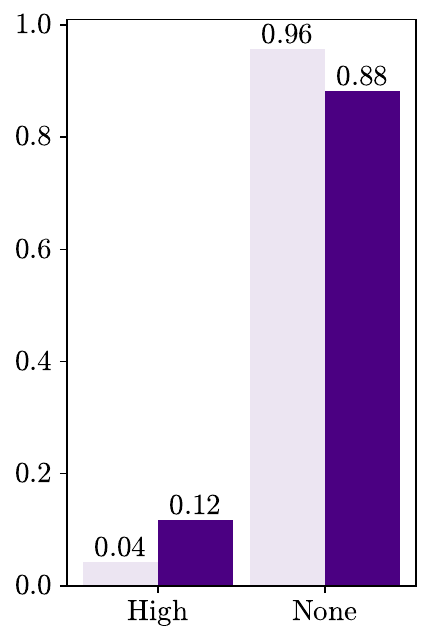}};
      \node[above=of img, node distance=0cm, rotate=0, anchor=center,yshift=-0.8cm,xshift=0.2cm] {\small{Merit}};
     \end{tikzpicture}
    \end{minipage} \\
    \footnotesize{p-value for $\chi^2$ test: \textbf{4.61e-11}} & \footnotesize{p-value for $\chi^2$ test: 0.052} & \centering\footnotesize{p-value for $\chi^2$ test: \textbf{4.22e-5}}
    \end{tabular}
    \caption{Distributions of tier, severity, and merit for all bugs received in the treated program before and after the reward change in July, 2024. Results for $\chi^2$ test of for distribution shift for each type variable are shown below each plot. The null hypothesis is that the category distribution is the same before and after the reward change. P-values $< 0.05$ are in bold.
    }
    \label{fig:type_hists}
\end{figure*}
\endgroup

\textbf{Observed changes in quantity for high-value bugs.}
Having shown that the distribution of bug types shifted, we now turn our attention towards 
changes in total counts per month and elasticities for high-value types. After all, the goal of the reward change was specifically to attract more high-value submissions.
Figure \ref{fig:highvalue_combined} shows the bug counts per month for Tier 0 bugs, bugs with Severity $\geq$ High, and High Merit bugs.
%(though we include all types in the Appendix). 
We observe particularly stark increases in bugs per month for Tier 0 and High Merit bugs, whereas a similar increase does not occur in the untreated programs.
% shows increases in bugs found per month for each of the high-value types for tier, severity, and merit. 
% The data contains Tier annotations starting in 2023, so we consider $\mathcal{T}$ to contain all months from January, 2023 through December, 2024. Figure \ref{fig:tier0_bpm} shows the number of Tier 0 bugs per month received by the treated program (similar figures for the rest of the tiers are given in the Appendix). 
Table \ref{tab:quantity_high_value} shows 
% the observed mean rate change and regression discontinuity design estimates for these high-value types. We observe 
that there is significant growth in the mean bug counts per month $\Delta \bar{Y}_x$ for all of these high-value types.\footnote{Aside from these high-value types, we also plot $\Delta \bar{Y}_x$ for all other types in the Appendix, though these all showed lower levels of significance and are of lower policy interest according to domain experts.} In the untreated programs, we do not observe statistically significant changes in $\Delta \bar{Y}_x$ in \textit{any} bug types.
We also observe low p-values for the robust $\hat{\tau}^{\text{RDD}}$ estimates for High Merit and Tier 0 bugs, which was not the case for all bugs in aggregate. This is suggestive of the reward change impacting high-value bugs more than all bugs in aggregate, which tracks with the fact that rewards increased the most for Tier 0 bugs. The significance of the $\hat{\tau}^{\text{RDD}}$ estimate for Tier 0 bugs is the closest to causal evidence in this section, although a full causal interpretation is limited by the local continuity assumption, and should be interpreted carefully against the possibility of external confounding events.

\begingroup
\setlength{\tabcolsep}{-2pt} 
\begin{figure*}[!ht]
    \centering
    \begin{tabular}{ccc}
    \small{\underline{Tier 0}} & \small{\underline{Severity $\geq$ High}} & \small{\underline{High Merit}} \\
    % \scriptsize{\underline{Bug counts in treated vs. untreated programs}} & \scriptsize{\underline{Regressions for treated program}} & \scriptsize{\underline{Gaps with permutation placebos}} \\
    \begin{minipage}{0.33\textwidth}
    \centering
    \begin{tikzpicture}
      \node (img)  {\includegraphics[scale=0.29]{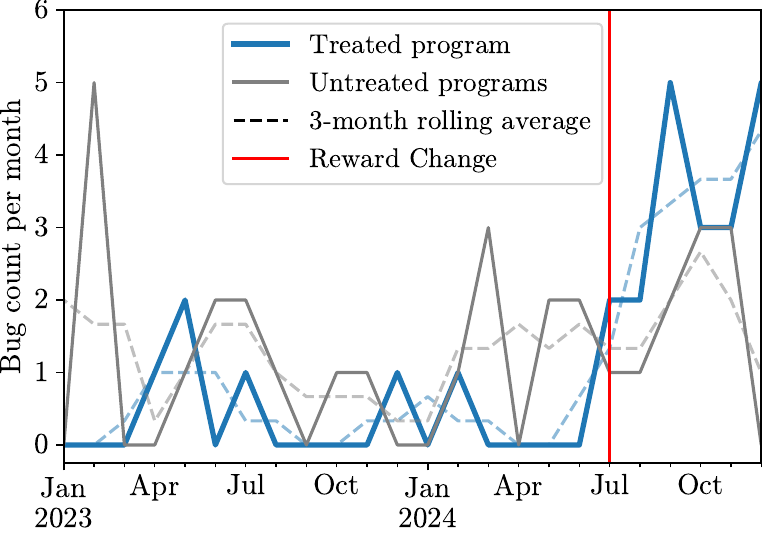}};
      \node[above=of img, node distance=0cm, rotate=0, anchor=center,yshift=-1cm,xshift=0.2cm] {\scriptsize{Bug counts}};
     \end{tikzpicture}
    \end{minipage} & 
    \begin{minipage}{0.33\textwidth}
    \centering
    \begin{tikzpicture}
      \node (img)  {\includegraphics[scale=0.29]{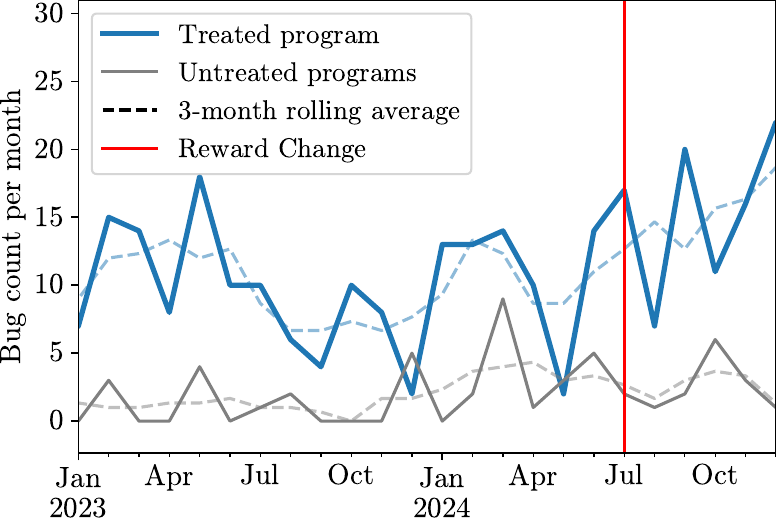}};
      \node[above=of img, node distance=0cm, rotate=0, anchor=center,yshift=-0.91cm,xshift=0.2cm] {\scriptsize{Bug counts}};
     \end{tikzpicture}
    \end{minipage}& 
    \begin{minipage}{0.33\textwidth}
    \centering
    \begin{tikzpicture}
      \node (img)  {\includegraphics[scale=0.29]{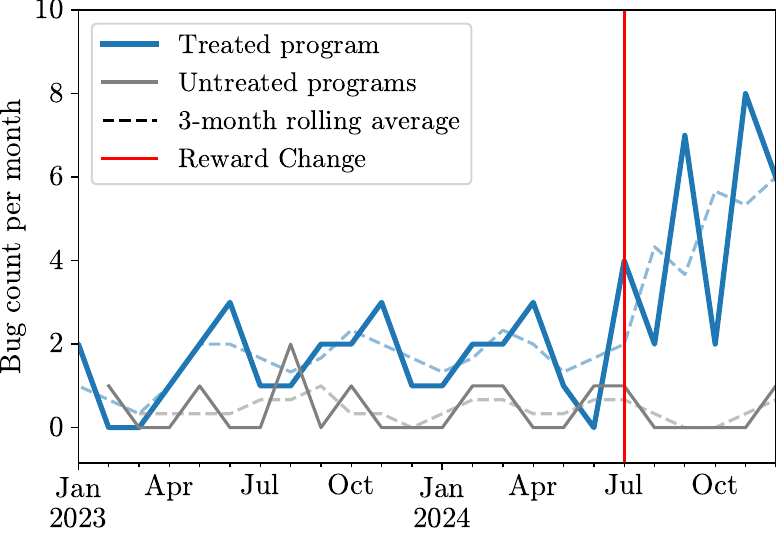}};
      \node[above=of img, node distance=0cm, rotate=0, anchor=center,yshift=-1cm,xshift=0.2cm] {\scriptsize{Bug counts}};
     \end{tikzpicture}
    \end{minipage} \\
    \begin{minipage}{0.33\textwidth}
    \centering
    \begin{tikzpicture}
      \node (img)  {\includegraphics[clip, trim={0 5pt 0 7.1pt}, scale=0.29]{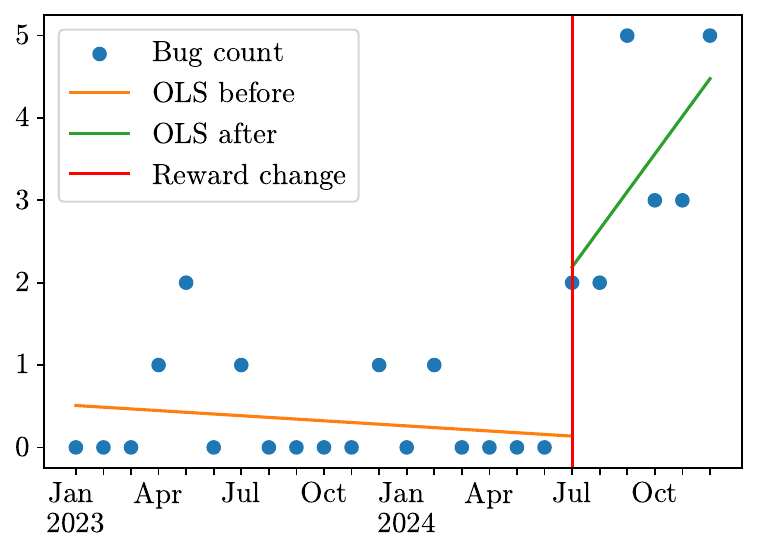}};
      \node[above=of img, node distance=0cm, rotate=0, anchor=center,yshift=-0.91cm,xshift=0cm] {\scriptsize{Regressions for treated program}};
    \end{tikzpicture}
    \end{minipage} & \begin{minipage}{0.33\textwidth}
    \centering
    \begin{tikzpicture}
      \node (img)  {\includegraphics[clip, trim={0 5pt 0 7.1pt}, scale=0.29]{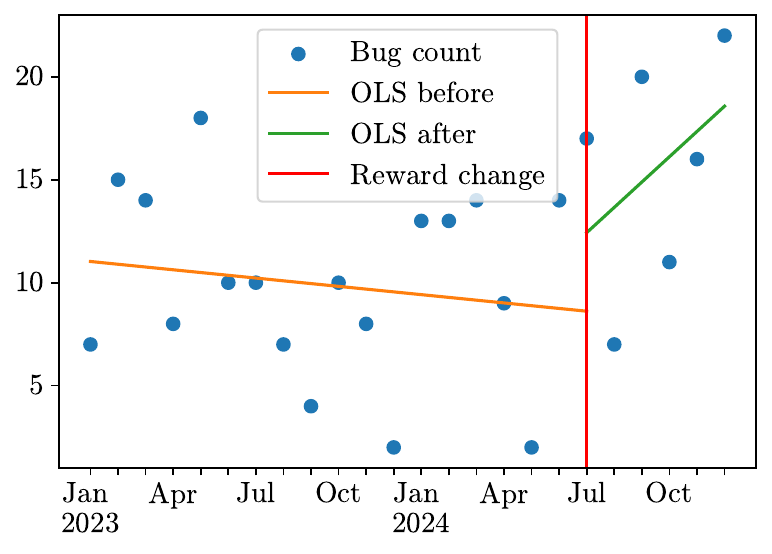}};
      \node[above=of img, node distance=0cm, rotate=0, anchor=center,yshift=-0.91cm,xshift=0cm] {\scriptsize{Regressions for treated program}};
    \end{tikzpicture}
    \end{minipage}& \begin{minipage}{0.33\textwidth}
    \centering
    \begin{tikzpicture}
      \node (img)  {\includegraphics[clip, trim={0 5pt 0 7.1pt}, scale=0.29]{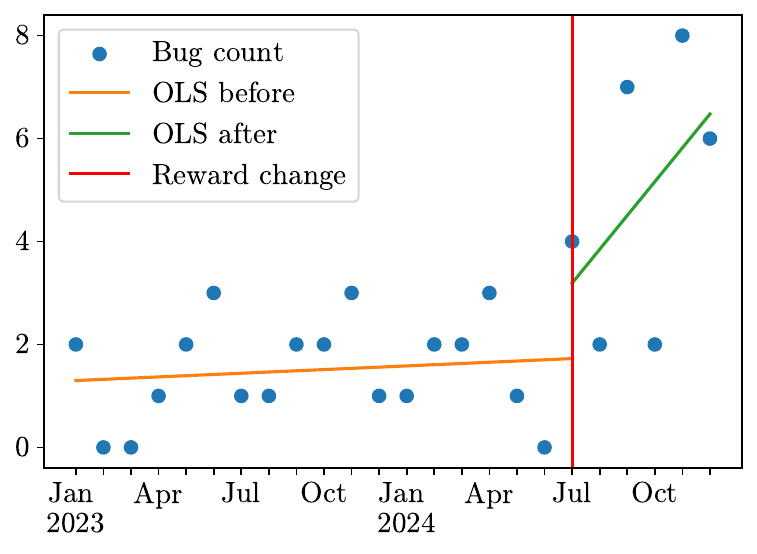}};
      \node[above=of img, node distance=0cm, rotate=0, anchor=center,yshift=-0.91cm,xshift=0cm] {\scriptsize{Regressions for treated program}};
    \end{tikzpicture}
    \end{minipage}
    \end{tabular}
    \caption{
    % Visualizations of bug counts and regressions for all bugs in the treated and untreated programs. 
      \textit{(Top row)} Bugs per month received in the treated and untreated programs for each of the high-value types. \textit{(Bottom row)} OLS regression fits in the treated program before and after the reward increase for each of the high-value types.}
    \label{fig:highvalue_combined}
\end{figure*}
\endgroup

\begin{table*}[!ht]
\setlength{\tabcolsep}{3pt} 
    \centering
    
    \caption{Quantity change statistics for high-value bugs in the treated program.}
    \begin{tabular}{c|c|c|c|c|c|c}
    \toprule
        & \multicolumn{2}{c|}{Tier 0} & \multicolumn{2}{c|}{Severity $\geq$ High} & \multicolumn{2}{c}{High Merit} \\
        \cline{2-7}
        Statistic & Value & \small{p-value} & Value & \small{p-value} & Value & \small{p-value} \\
        \midrule
        $\Delta \bar{Y}_x$ & \textbf{3.00}  & \textbf{1.30e-7} & \textbf{5.45} & \textbf{0.023} & \textbf{3.17} & \textbf{1.55e-4}  \\
        % $\hat{\tau}^{\text{DiD}}_x$ & \textbf{2.15}  & \textbf{0.005} & 4.81 & 0.069 & \textbf{4.28} & \textbf{0.0001} \\
        % OLD RDD VALUES
        % $\hat{\tau}^{\text{RDD}}_x$ & \textbf{2.05} &  \textbf{0.005} & 3.81 & 0.378 & 0.79 & 0.569\\
        % $\hat{\tau}^{\text{RKD}}_x$ & \textbf{0.47} & \textbf{0.017} & 1.36 & 0.261 & \textbf{1.23} & \textbf{0.004}\\
        % ROBUST RDD VALUES
        $\hat{\tau}^{\text{RDD}}_x$ & \textbf{2.34} &  \textbf{0.028} & 6.33 & 0.418 & 3.95 & 0.067\\
        $\hat{\tau}^{\text{RKD}}_x$ & 1.04 & 0.377 & -1.98 & 0.671 & 1.07 & 0.548\\
        Chow test & N/A & \textbf{2.16e-5} & N/A & 0.135 & N/A & \textbf{0.029}\\
    \bottomrule
    \end{tabular}
    
    \label{tab:quantity_high_value}
\end{table*}

\textbf{Observed elasticities for high-value bugs.} 
While it seems intuitively obvious that a higher reward increase should lead to a higher increase in bug quantities, the question remains of how much, exactly, a marginal increase in reward affects the outcome. Notably, we find in Table \ref{tab:elasticities} that the elasticity estimates of Tier 0 and High Merit bugs are both greater than 1. That is, the findings of Tier 0 and High Merit bugs doubled after the rewards doubled.
% It is perhaps expected that we would observe a higher quantity change for high-value types as these are receiving a higher reward change; however, it is surprising to see a higher \textit{elasticity} for the high-value bug types. 
This suggests that there could be more potential among researchers to divert attention towards finding high-value bugs, whereas researchers may be finding close to as much as they can find of the low hanging fruit. Again, a full causal interpretation should be made with caution against the possibility of external confounding events.

% Thus, we now consider the elasticities for these high-value types.
% Table \ref{tab:elasticities} shows estimated elasticities for the changes in quantity for each of the high-value bug types. 
% All of these elasticity estimates are higher than the elasticity for the overall bugs, but most significantly, the elasticity estimates for Tier 0 and High Merit bugs are greater than 1. It is perhaps expected that we would observe a higher quantity change for high-value types as these are receiving a higher reward change; however, it is surprising to see a higher \textit{elasticity} for the high-value bug types. This indicates that perhaps there is more potential among researchers to divert attention towards finding high-value bug types, but researchers are finding closer to as much as they can find of the low hanging fruit.

\begin{table*}[!h]
\setlength{\tabcolsep}{3pt} 
    \centering
    
    \caption{Elasticity estimates for high-value bug types.
    }
    \begin{tabular}{c|c|c|c|c|c|c|c}
    \toprule
         Bug type & $\bar{Y}_0^x$ & $\bar{Y}_1^x$ & $\%\Delta \bar{Y}_x$ & $\bar{R}_0^x$ & $\bar{R}_1^x$ & $\%\Delta \bar{R}_x$ & $\hat{\eta}_x$ \\
         \midrule
         Tier 0 & $0.33$ & $3.33$ & $900.00\%$ & $\$7,416.66$ &  $\$16,625.55$ & $124.16\%$ & $7.24$\tiny{[CI: $(2.20, 32.44)$]} \\
         Severity $\geq$ High & $9.88$ & $15.33$ & $55.05\%$ & $\$2,291.80$ &  $\$7,558.48$ & $229.80\%$ & $0.24$\tiny{[CI: $(0.08, 0.58)$]} \\
         High Merit & $1.50$ & $4.67$ & $211.11\%$ & $\$5,063.82$ &  $\$13,016.69$ & $157.05\%$ & $1.34$\tiny{[CI: $(0.36, 5.39)$]} \\
    \bottomrule
    \end{tabular}
    \label{tab:elasticities}
\end{table*}

\section{Who is driving the increases in found bugs?}\label{sec:top_researchers}

Having shown that the reward change resulted in more high-value bugs being found, a natural follow-up question is, \textit{who is behind this?} Is there an influx of new researchers entering the treated program after the reward change? Are existing researchers becoming more productive? Or perhaps some mix of both? 
To understand this, we dive deeply into disentangling the output from \textit{veteran researchers} (i.e., all researchers who had ever participated in the program before the reward change) vs. \textit{new researchers} (i.e., those who submit a report for the first time after the reward change).\footnote{Technically ``new'' here means new since 2018.} Studies in labor economics often aim to distinguish between these effects, using the terminology ``intensive margin'' to refer to production by existing labor and resources, and ``extensive margin'' to refer to effects from new entrants. We structure this section in two parts: 

\begin{enumerate}[topsep=0pt,itemsep=0pt]
    \item First, we focus on the outputs from veteran researchers. We seek to understand whether veteran researchers are increasing their production, or driving the increase in bugs found in the treated program. %\textit{Are veteran researchers driving the increase in bugs found in the treated program?} 
    As a preview, we find that veteran researchers play a significant role in the increases in high-value bugs. At the same time, a large portion of the increases in overall bug counts can be attributed to new researchers after the reward change.
    \item Second, we focus on the new researchers attracted after the reward change. The goal is to disentangle quality from quantity: did the reward change attract \textit{more} new researchers or \textit{more productive} new researchers? Our findings are somewhat subtle: instead of a general influx in the quantity of new researchers, we find that the reward change attracted a relatively small number of highly productive researchers. 
\end{enumerate}

Taken as a whole, this breakdown gives insight into the effects on attraction and retention of an increase in rewards, which can be important for policy decisions in this and other bug bounty programs. The finding that the reward increase attracted new highly productive researchers is particularly important in a competitive environment, where various programs may compete to attract the effort of top researchers.

\subsection{Outputs from veteran researchers}

\begin{figure}[!ht]
    \centering
    \begin{minipage}{0.48\textwidth}
    \centering
    \begin{tikzpicture}
      \node (img)  {\includegraphics[scale=0.35]{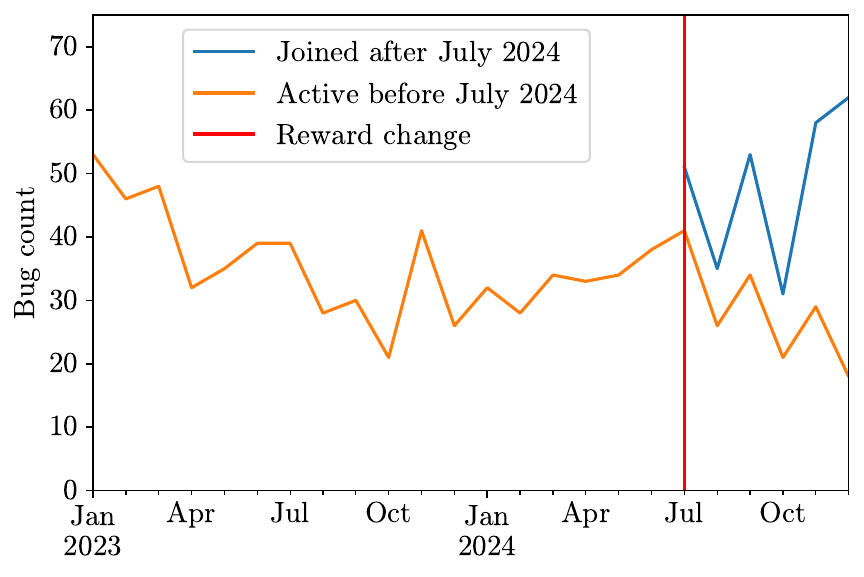}};
      \node[above=of img, node distance=0cm, rotate=0, anchor=center,yshift=-0.9cm,xshift=0cm] {\scriptsize{Breakdown of bugs between veteran and new researchers}};
     \end{tikzpicture}
    \end{minipage} \begin{minipage}{0.48\textwidth}
    \centering
    \begin{tikzpicture}
      \node (img)  {\includegraphics[scale=0.35]{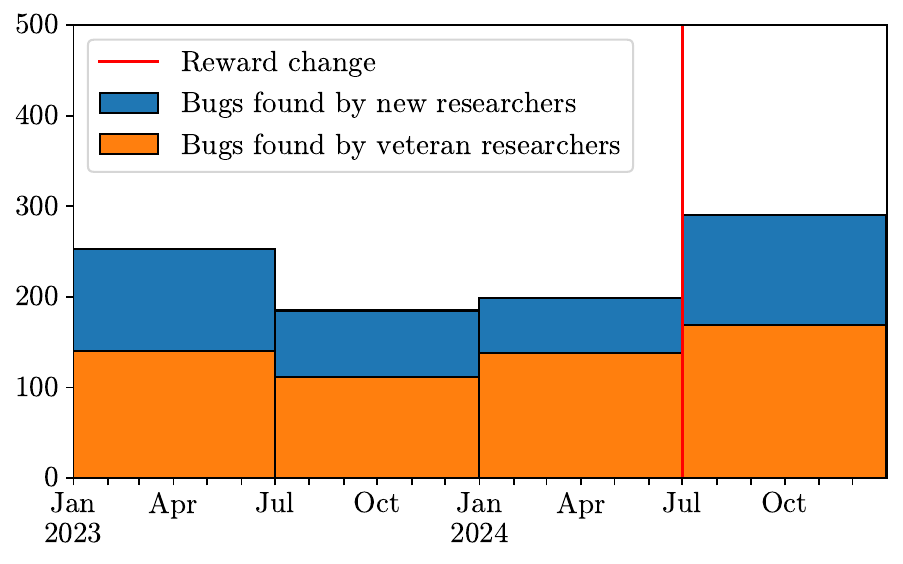}};
      \node[above=of img, node distance=0cm, rotate=0, anchor=center,yshift=-0.9cm,xshift=0cm] {\scriptsize{Shares of bugs found by veteran vs. new researchers}};
     \end{tikzpicture}
    \end{minipage} 
    \caption{\textit{(Left)} Breakdown of bugs found by veteran researchers and new researchers after the reward change in the treated program. The \textcolor{orange}{orange} line tracks bugs found by veteran researchers who were part of the program before the reward change (intensive margin). The \textcolor[rgb]{0.12,0.47,0.71}{blue} line includes bugs found by new researchers who joined after the reward change (extensive margin). \textit{(Right)} Shares of bugs found by veteran researchers vs. new researchers in each six-month window. Veteran researchers are defined as anyone who has ever submitted a report to the program before the start of the window, and new researchers are those who submitted for the first time during the window.}
    \label{fig:margins_2023-2024}
\end{figure}

We begin by disentangling the bug counts in terms of those found by veteran researchers and new researchers. A primary goal of this section is to answer the question, \textit{To what extent are veteran researchers driving the increase in bugs found after the reward change?}
% In economic terms, we refer to the contribution of veteran researchers as the intensive margin, and the contribution of all researchers, including new entrants, as the extensive margin. 
Figure \ref{fig:margins_2023-2024} shows a breakdown of Figure \ref{fig:all_timeseries} into bugs found by veteran researchers and the bugs found by new researchers joining after the reward change. 
% The orange line illustrates the intensive margin, or the change in bugs found by existing participants, and the blue line illustrates the extensive margin, or the additional bugs found by new entrants.
Overall, there appear to be growing contributions from new researchers after the reward change. 
An important subtlety in this analysis is that there is always churn in researcher participation, and perhaps some drop in production of veteran researchers is to be expected as researchers leave or exhaust their resources. Figure \ref{fig:margins_2023-2024} \textit{(right)} shows that the \textit{share} of total bugs found by veteran researchers does not seem to differ significantly from previous time periods.
% Thus, Figure \ref{fig:margins_2023-2024} does not rule out that the reward change had some impact on veteran researchers: even if the production from veteran researchers didn't strictly increase, did the production decrease \textit{less} compared to previous arbitrary months when there wasn't a reward change? 
% To answer this question, note that in any time window, there is always a split between the contribution of veteran researchers, and new researchers joining for the first time. 
% To see if the split after the reward change is unusual, Figure \ref{fig:margins_2023-2024} \textit{(right)} illustrates the share of bugs found by veteran researchers and the share of bugs found by new researchers in each 6-month window before and after the reward change. The share of veteran contribution after the reward change does not seem to be unusually high, and the level of ``drop-off'' in production from veteran researchers relative to the full production in the previous period does not seem to be unusual after the reward change.

% \begin{figure}[!ht]
%     \centering
%     \begin{minipage}{0.5\textwidth}
%     \centering
%     \begin{tikzpicture}
%       \node (img)  {\includegraphics[scale=0.35]{figures/shares.pdf}};
%       \node[above=of img, node distance=0cm, rotate=0, anchor=center,yshift=-0.9cm,xshift=0cm] {\scriptsize{Shares of bugs found by veteran vs. new researchers}};
%      \end{tikzpicture}
%     \end{minipage} 
%     \caption{Shares of bugs found by veteran researchers vs. new researchers in each six-month window. In each six-month window, veteran researchers are defined as anyone who has ever submitted a report to the program before the start of the window, and new researchers are those who submitted for the first time during the six-month window.}
%     \label{fig:shares}
% \end{figure}

% The share of veteran contribution after the reward change does not seem to be unusually high, and the level of ``drop-off'' in production from veteran researchers relative to the full production in the previous period does not seem to be unusual after the reward change. Instead, there appears to be a higher proportion of bugs found by new researchers after the reward change.  This proportion is higher than both the prior 6 months, and the same time window in 2023. 
% %This supports the idea that new researchers are driving the increase in overall bug counts after the reward change.

Things get more interesting when we consider high-value bugs. Figure \ref{fig:shares-types} gives a breakdown of Figure \ref{fig:margins_2023-2024} \textit{(right)} into the previously defined high-value bug categories. There are two notable findings from this breakdown. First, for both Tier 0 and High Merit bugs, the contribution of veteran researchers significantly increased after the reward change. This suggests that the reward change has effectively \textit{redirected} veteran researcher efforts towards high-value bugs. Second, the contribution of new researchers is higher after the reward change for all high-value bug types. This is especially significant in an environment where multiple bug bounty programs are competing for the attention of the same pool of researchers.
% This holds both in terms of absolute bug counts and in terms of the proportion relative to bugs found by veteran researchers.

\begingroup
\setlength{\tabcolsep}{0pt}
\begin{figure*}[!ht]
    \centering
    \begin{tabular}{ccc}
    % \small{Tier 0} & \small{Severity $\geq$ High} & \small{High Merit} \\
    \begin{minipage}{0.33\textwidth}
    \begin{tikzpicture}
      \node (img)  {\includegraphics[scale=0.26]{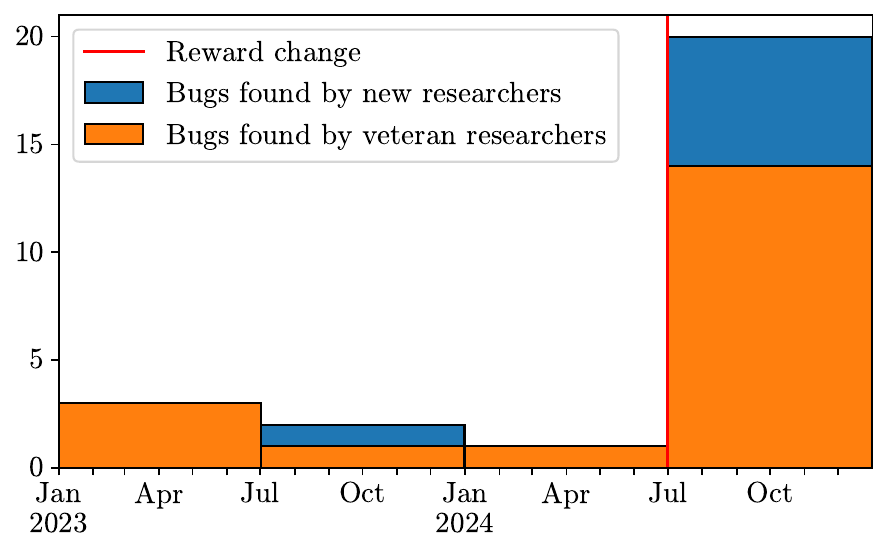}};
      \node[above=of img, node distance=0cm, rotate=0, anchor=center,yshift=-0.9cm,xshift=0cm, text width=\textwidth, align=center] {\scriptsize{Tier 0}};
     \end{tikzpicture}
    \end{minipage} & 
    \begin{minipage}{0.33\textwidth}
    \begin{tikzpicture}
      \node (img)  {\includegraphics[scale=0.26]{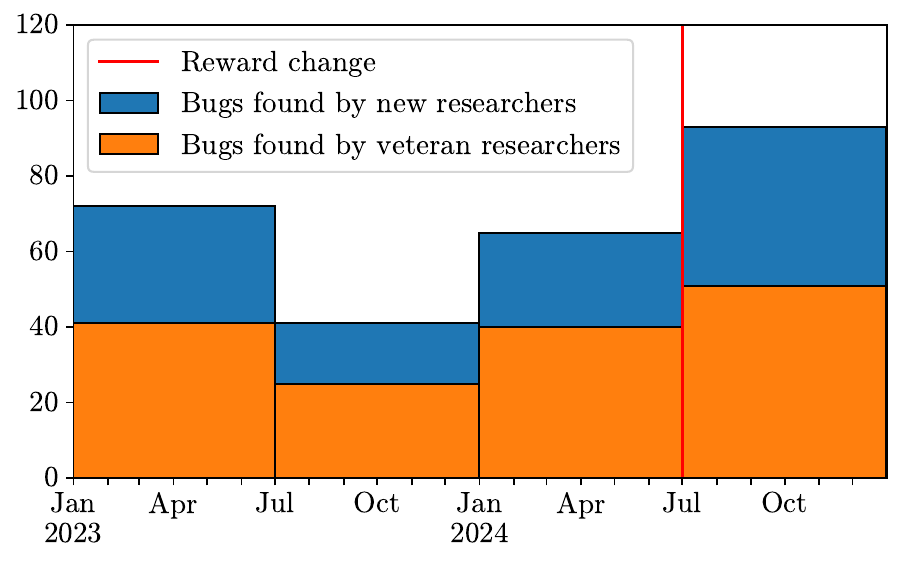}};
      \node[above=of img, node distance=0cm, rotate=0, anchor=center,yshift=-0.9cm,xshift=0cm, text width=\textwidth, align=center] {\scriptsize{Severity $\geq$ High}};
    \end{tikzpicture}
    \end{minipage} &
    \begin{minipage}{0.33\textwidth}
    \begin{tikzpicture}
      \node (img)  {\includegraphics[scale=0.26]{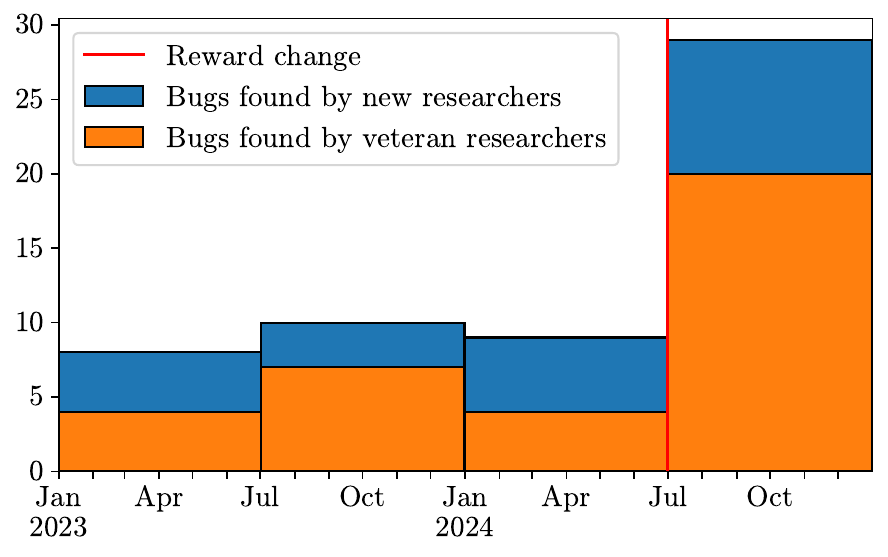}};
      \node[above=of img, node distance=0cm, rotate=0, anchor=center,yshift=-0.9cm,xshift=0cm, text width=\textwidth, align=center] {\scriptsize{High Merit}};
     \end{tikzpicture}
    \end{minipage} 
    \end{tabular}
    \caption{Shares of bugs found by veteran researchers vs. new researchers in each six-month window, separated by high-value types.
    }
    \label{fig:shares-types}
\end{figure*}
\endgroup

\subsection{Analysis of new researchers}
Having shown that new researchers play a significant role in the increase in bug counts after the reward change, we next break this down more finely to better understand whether the reward change has attracted more new researchers, whether researchers are transferring between programs, or whether the new researchers joining are somehow more productive than before. As a preview, we find that the reward change attracted a \textit{new, relatively small cohort of highly productive researchers}. 

\begin{figure}[!ht]
    \centering
    \begin{minipage}{0.45\textwidth}
    \centering
    \begin{tikzpicture}
      \node (img)  {\includegraphics[scale=0.35]{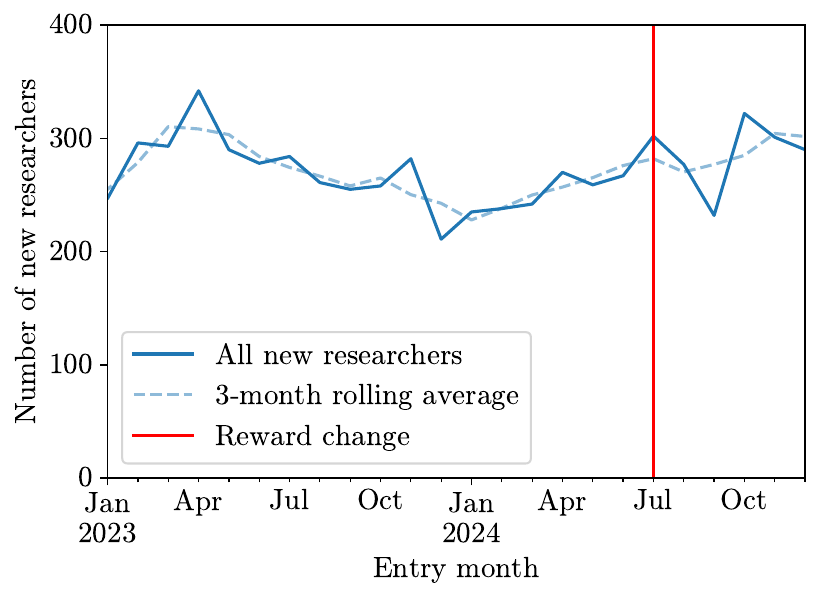}};
      \node[above=of img, node distance=0cm, rotate=0, anchor=center,yshift=-0.9cm,xshift=0cm] {\scriptsize{Number of new researchers}};
     \end{tikzpicture}
    \end{minipage} 
    \caption{New researchers per month who submit a report for the first time to the treated program.}
    \label{fig:new}
\end{figure}

\textbf{Total new researchers.} We first show that the raw number of new researchers entering the program does not appear to have significantly increased after the reward change. In a given month, we define a new researcher as someone who submits a report for the first time to the program in that month. Figure \ref{fig:new} shows that the number of new researchers entering the treated program each month stayed relatively constant.
\footnote{The same is true if we filter only for new ``successful'' researchers, see Figure \ref{fig:new_successful} in the Appendix.}

\textbf{Detailed productivity breakdown.} To fully explain the increase in bugs found by new researchers, we analyze the productivity of all new researchers in detail. We measure productivity per researcher as the number of bugs found in their first six-months, so that the counts are roughly comparable regardless of when the researcher first entered the program. Figure \ref{fig:before-after-hists-all} shows a full breakdown of productivity per researcher for new researchers entering before and after the reward change. 
Importantly, the increase in bugs found by new researchers after the reward change is not driven by a single outlier, but rather a \textit{group} of researchers entering after the reward change, who all have fairly high productivity compared to those who joined in the period before the reward change. The length of the ``tail'' of the distribution does not show much change, but there are more highly productive new researchers after the reward change. None of these new highly productive researchers after the reward change were transfers from other Google programs.

Breaking this down into high-value bug types, Figure \ref{fig:before-after-hists-types} shows that the reward change attracted new researchers who were productive at finding high-value bugs, and this was again not limited to a single outlier. Compared to the period before the reward change, the new researchers who joined after the reward change found more high-value bugs. A few of these researchers who found Tier 0 and High Merit bugs were transfers, but the majority were not.

\begingroup
\begin{figure*}[!h]
    \centering
    \begin{tabular}{c}
    \begin{minipage}{0.5\textwidth}
    \centering
    \begin{tikzpicture}
      \node (img)  {\includegraphics[clip, trim={7pt 5pt 0 7pt}, scale=0.4]{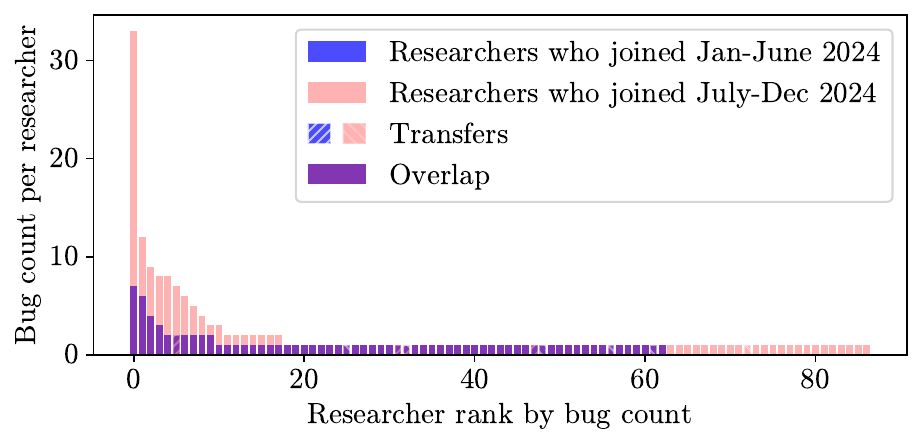}};
      \node[above=of img, node distance=0cm, rotate=0, anchor=center,yshift=-0.9cm,xshift=0cm, text width=\textwidth, align=center] {\footnotesize{First six-month bug counts per researcher}};
      % \\ joining before and after the reward change.}};
     \end{tikzpicture}
    \end{minipage} \\
    \end{tabular}
    \caption{First six-month bug counts for new researchers joining before and after the reward change. Each bar represents a single researcher's bug count in their first six months. Counts are sorted in decreasing order (producing a ``researcher rank'' on the horizontal axis).
    }
    \label{fig:before-after-hists-all}
\end{figure*}
\endgroup

\begingroup
\setlength{\tabcolsep}{2pt}
\begin{figure*}[!h]
    \centering
    \begin{tabular}{ccc}
    % \footnotesize{Tier 0} & \footnotesize{Severity $\geq$ High} & \footnotesize{High Merit} \\
    \begin{minipage}{0.32\textwidth}
    \begin{tikzpicture}
      \node (img)  {\includegraphics[clip, trim={7pt 5pt 0 7pt}, scale=0.3]{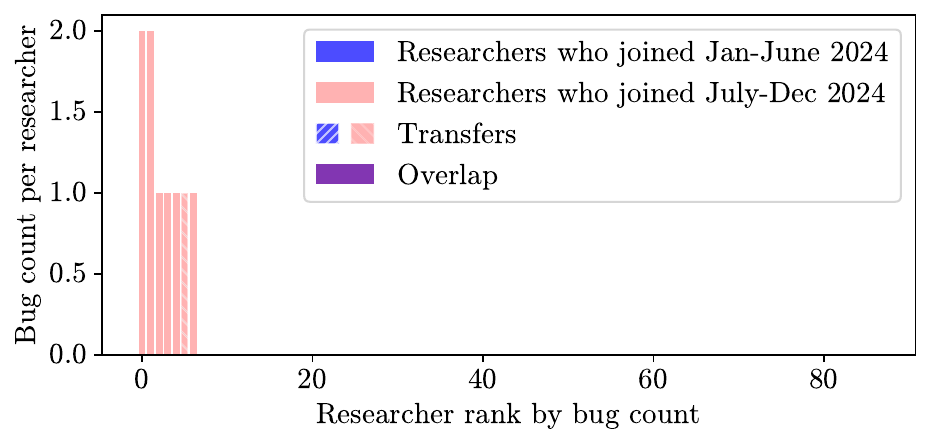}};
      \node[above=of img, node distance=0cm, rotate=0, anchor=center,yshift=-0.9cm,xshift=0cm, text width=\textwidth, align=center] {\scriptsize{Tier 0}};
     \end{tikzpicture}
    \end{minipage} & 
    \begin{minipage}{0.32\textwidth}
    \begin{tikzpicture}
      \node (img)  {\includegraphics[clip, trim={7pt 5pt 7pt 7pt}, scale=0.3]{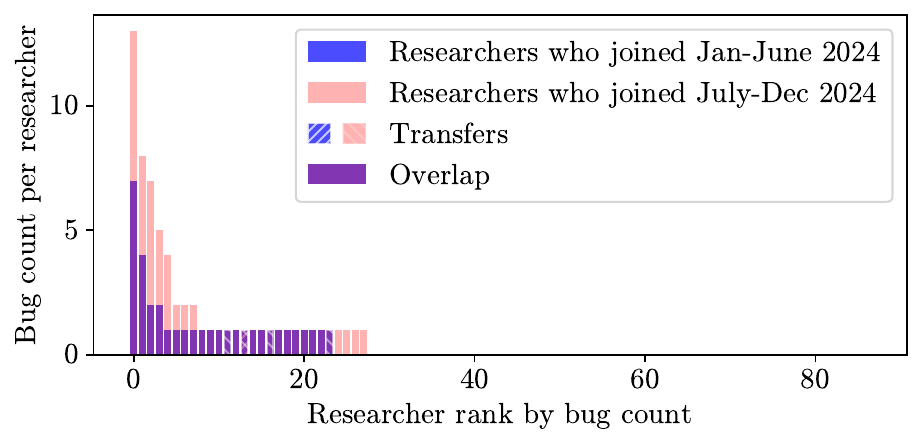}};
      \node[above=of img, node distance=0cm, rotate=0, anchor=center,yshift=-0.9cm,xshift=0cm, text width=\textwidth, align=center] {\scriptsize{Severity $\geq$ High}};
    \end{tikzpicture}
    \end{minipage} &
    \begin{minipage}{0.32\textwidth}
    \begin{tikzpicture}
      \node (img)  {\includegraphics[clip, trim={7pt 5pt 0 7pt}, scale=0.3]{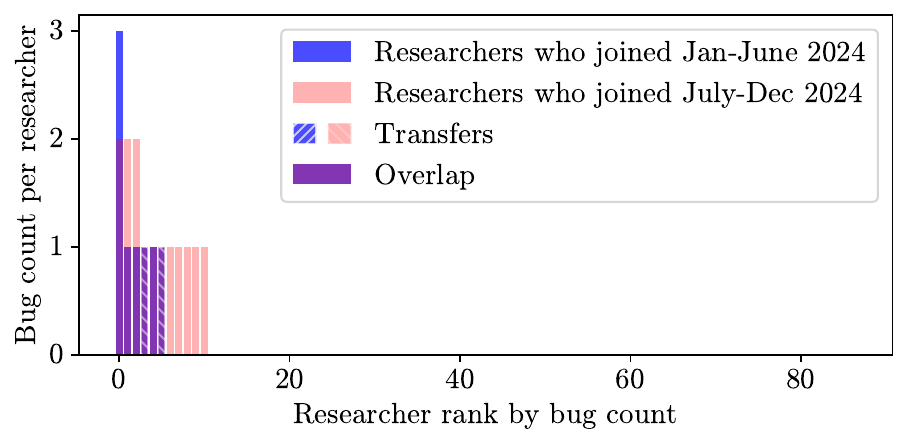}};
      \node[above=of img, node distance=0cm, rotate=0, anchor=center,yshift=-0.9cm,xshift=0cm, text width=\textwidth, align=center] {\scriptsize{High Merit}};
     \end{tikzpicture}
    \end{minipage} 
    \end{tabular}
    \caption{First six-month bug counts for new researchers joining before and after the reward change, separated by high-value types. Each bar represents a single researcher's bug count for the high-value bug type in their first six months.
    % First six-month bug counts for all researchers who joined before and after the price change. Each bar represents a single researcher's bug count in their first six months in the program. The \textit{top} plot includes all bugs, and the \textit{bottom} three plots show only high-value bugs found. 
    }
    \label{fig:before-after-hists-types}
\end{figure*}
\endgroup

Taken together, our analysis of new researchers show that the reward change managed to attract a small number of highly productive researchers. The new researchers arriving after the reward change were more productive in their first six months than new researchers arriving in previous periods. There could be several partial explanations for this. \textit{(i)} It could be that the higher rewards attracted different types of new researchers compared to before, such as those with more experience, or more developed tools and skills. \textit{(ii)} It could also be that new researchers arriving after the reward change were willing to put in more effort than new researchers arriving before, due to the higher rewards. While the counterfactual does not exist, it could be that if the same researcher were to join after the reward change vs. before, their initial productivity would be higher if they joined after. \textit{(iii)} A third partial explanation could be that the reward change attracted some highly-skilled researchers from other bug bounty programs.

Fully disentangling these partial explanations is beyond the scope of the data analyzed here, as there is no additional information on researcher types or participation in other programs. The fact that cross-program substitution within Google was sparse 
% (more detail in Appendix \ref{app:substitution}) 
suggests that the new incoming productive researchers could be coming from working on similar software domains.
Ultimately, from a policy standpoint, these results suggest that increasing rewards is a viable way to attract participation from highly-skilled researchers, but may not in itself grow the volume of participation from new inexperienced researchers. However, highly-skilled researchers also tend find high-value bugs.
\section{Discussion and Conclusions}\label{sec:conclusions}

Our analysis of Google VRP data provides a number of insights into the responsiveness of outcomes and labor to changes in bug bounty reward incentives. These insights indicate effectiveness of the bug bounty program as a whole, and also lay the groundwork for developing future improvements in the design of bug bounty programs for better security outcomes.
%mechanisms, and third-party information elicitation mechanisms more generally.
Most notably from a security standpoint, we observe statistically significant increases in the reporting of high-value bugs, especially in the highest impact tiers and high merit submissions. The high merit submissions are of particular interest because these often indicate exceptional quality and novelty, which tend to have a higher security impact since they're often easier for internal security engineers to turn into actionable changes. Interestingly, elasticity estimates show that the response to the reward change was significantly \textit{higher} for high-value bugs. This suggests further room for the program to grow its high-value outcomes through additional reward increases. 
% The significantly lower elasticity from lower impact bugs indicates
Our distributional analysis indicated possible substitution effects from medium-value bugs, which could also be a refinement to the program, as these bugs still require resources to triage. 
These findings lay the groundwork for developing a further optimized reward scheme in the future to elicit a more beneficial set of bugs.

% From a labor standpoint, we also observe a higher elasticity among top researchers compared to the more occasional single-bug submitters -- in fact, the productivity of single-bug researchers stayed remarkably constant throughout the observation period. Top researchers tended to not only find higher value bugs overall, but also responded more strongly to incentives.

% Broadly speaking, these results bolster a story about two types of security researchers: those who find bugs in the process of using Google products for other reasons like their employment, and those for whom finding bugs is a significant activity (similar to prior models that differentiate ``expert'' researchers from ``non-expert'' researchers \citep{gal2024merchants}).  We call these incidental and professional researchers, respectively. Incidental researchers could be motivated by factors other than the reward amounts, perhaps submitting as a hobby, or submitting a bug that they come across in their regular work or usage activities. Payments need to be sufficient to induce incidental researchers to file bugs, which take time to replicate and document, but beyond that, increases in payments have little to no effect on their bug reporting activity. It could also be that the low-hanging fruit found by incidental researchers is already relatively saturated at the current reward amounts, whereas there may still be harder-to-find bugs available if the effort were adequately compensated. 

From a labor standpoint, the analysis of the effects of the behavior of veteran researchers and new researchers gives a view into the effects of the reward increase on the retention, redirection, and attraction of researchers. Notably, veteran researchers were a primary driver for the increase in high-value bugs, but not necessarily for all bugs. On the other hand, the reward change clearly attracted a small number of new highly productive researchers, who contributed to both an increase in overall bug counts and increases in high-value bug counts.

\textbf{A tale of two researcher types.} Connecting these findings to theoretical models, our breakdown of researcher productivity aligns with a story of two types of security researchers: those for whom finding bugs is a significant activity, and those who find bugs in the process of using Google products for other incidental reasons (similar to prior theoretical models that differentiate ``expert'' from ``non-expert'' researchers \citep{gal2024merchants}).  We call these professional and incidental researchers, respectively.

Incidental researchers could be motivated by factors other than the reward amounts, perhaps submitting as a hobby, or submitting a bug that they come across in their regular usage. Payments need to be sufficient to induce incidental researchers to file bugs, which take time to replicate and document, but beyond that, increases in payments have little to no effect on their bug reporting activity. 
% It could also be that the low-hanging fruit found by incidental researchers is already relatively saturated at the current reward amounts.
%whereas there may still be harder-to-find bugs available if the effort were more highly compensated.
Given that such incidental researchers make up the majority of participants, this would track with the finding that the raw number of new researchers did not increase significantly after the reward change, nor was there much change to the ``tail'' of the productivity of new researchers entering before and after the reward change.

Professional researchers, on the other hand, view security research 
% (possibly not just through Google's bug bounty programs, but also through the programs run by and for other firms)
as a major income source and value reward changes more. For professionals already focused on Google products, the finding that veteran researchers found more high-value bugs after the reward change could indicate that these professional researchers always had the capacity to find such bugs, and the reward change now compensates them well enough for the higher effort it takes to search for Tier 0 bugs and put together high merit reports. For professionals focused on other platforms, the finding that the reward change attracted new highly productive researchers could suggest that these professionals may be attracted by the higher payments to switch platforms, resulting in the entry of new top researchers.

\textbf{Future work in an AI era.}
% More generally, this work lays a foundation for future empirical and theoretical work on the value and design of bug bounty programs, even as AI enters the changing landscape.
The emergence of AI use in cybersecurity \citep{mythos,bugbot,bigsleep} has already dramatically changed the bug bounty landscape. While this study focused on human effort in 2024 (before these AI bug hunting tools were released), the economic landscape of bug hunting will continue to shift in the upcoming years. In 2025, the widespread adoption of AI tools for writing bug reports led to an influx of ``slop'' reports \citep{deathbyathousandslops,aireports}, which dramatically increased the load on reviewers, sometimes to a critical degree. The landscape has continued to change as AI capabilities have improved in 2026. 
Going forward, two important questions stand out regarding the role of AI in bug bounty programs: \textit{(i)} To what extent does AI help bug hunters find and report high-value bugs more efficiently? \textit{(ii)} How effective are AI tools at supporting triage for those who run bug bounty programs? At a minimum, repeating the types of analysis in this work in similar programs could yield insights into both questions. The results in this work can serve as a necessary comparative baseline for any such future studies.

\section*{Acknowledgments}
We thank P. M. Aronow, Andrei Broder, Jon Gill, Christoph Kern, Ravi Kumar, Aranyak Mehta, James Mickens, and Martin Straka for the valuable feedback in the development of this work.

%
%\begin{acks}
%
%	The authors would like to thank Dr. Maura Turolla of Telecom
%	Italia for providing specifications about the application scenario.
%
%	The work is supported by the \grantsponsor{GS501100001809}{National
%		Natural Science Foundation of
%		China}{http://dx.doi.org/10.13039/501100001809} under Grant
%	No.:~\grantnum{GS501100001809}{61273304\_a}
%	and~\grantnum[http://www.nnsf.cn/youngscientsts]{GS501100001809}{Young
%		Scientsts' Support Program}.
%
%
%\end{acks}

% Bibliography
\bibliography{references}
\newpage
\clearpage

% Appendix
\appendix
\section{Limitations} The methods and data in this work come with assumptions and complications, and we discuss gaps that future data sources and analyses may be able to help fill. 
We have shown statistically significant observational changes, but the estimates of causal effect rely on identification assumptions which are inevitably oversimplified relative to reality. One notable complication is the possibility of delayed effects of the reward increase: as high-value bugs can take months to find, we may not observe an immediate increase in bugs found the moment the reward change is deployed. Knowledge of the reward change could also take time to disseminate. 
% Furthermore, the actual realized rewards after bonuses and penalties differ from the amounts announced in the reward table.
Future work verifying our assumptions, or analyses that rely on different sets of assumptions, would be valuable. 

Our analysis has also been intentionally limited to short-term effects in the six months after the reward change. We would have liked to study longer term effects; however, more time passing came with more external confounding factors. The bug bounty landscape changed significantly in 2025, with a rise in popularity of AI tools for bug hunting \citep{aireports}, as well as new AI bug bounty programs and significant reward increases outside Google \citep{anthropic,openaiincrease,appleincrease,google_ai_bug_bounty}. Still, we observe persistent growth in bug counts into 2025 (see Appendix \ref{app:2025}), but it is hard to say how much of this is due to Google's reward increase, and how much is due to other external factors and program changes. 
% We include further plots into July, 2025 in Appendix \ref{app:2025}.
% We include further plots into July, 2025 in the Appendix; however, this data is affected by a rise in popularity of AI tools for bug hunting, as well as the creation of multiple new AI bug bounty programs. We observe persistent growth in bug counts into 2025, but it is hard to say how much of this is due to the reward increase, and how much is due to other external factors and program changes.

Another confounding factor that we have not directly addressed is the \textit{availability} of bugs in the system. As software systems are constantly evolving with new code and features being deployed, the number and types of bugs that \textit{exist} in the system also vary. While we are unaware of significant changes to the Google VRP's software targets in 2024, we do know that the data in 2025 is affected by the addition of new AI products as software targets for the AVRP.
% Future work that brings in data on the availability of bugs would be interesting.
%extremely interesting; however, we were not able to obtain such data for this study.

Finally, an important confounding factor that is worth significant future attention is the possible presence of exogenous forces on researcher behavior -- perhaps from reward increases at other companies, or exploit brokers and black market offers. Exploit broker reward offers tend to be orders of magnitude higher than bug bounty program offerings; however, exploit brokers also tend to only be interested in a much more select set of bug types. To our knowledge, there have not been any major reward changes in primary competing programs with the GAVRP in the study period, but we were not able to verify this for all possible outside programs or exploit brokers. We are also not aware of work that analyzes the substitutability of researcher effort between different programs. Future work that directly studies substitution across programs would be highly valuable, and it would be interesting to compare substitution effects with the elasticities found in this work.

\subsection{Additional future work}
More generally, this work lays a foundation for future empirical and theoretical work on the value and design of bug bounty programs, even as AI enters the changing landscape.

\textbf{External vs. internal outcomes in third-party programs.} 
% More generally, this work lays a foundation for future empirical and theoretical work on the value and design of third-party bug bounty programs.
%information elicitation mechanisms in firm operations across domains.
% Such work across domains is becoming increasingly important as third-party information elicitation mechanisms take a growing role in the training and evaluation of AI, via mechanisms like Reinforcement Learning with Human Feedback (RLHF) \citep{bai2022training}, crowdsourced data labeling \citep{zhang2016learning}, and red-teaming for AI safety \citep{feffer2024red}.
A major open question following from this work is whether the high-value bugs found from this bug bounty program actually \textit{differ} from bugs that would have been found through regular internal debugging processes. Such internal processes include a pipeline of better code development tools, better code review processes, and dedicated bug-finding by internal engineers or automated tools.
% Answering this question would require a comparison to firm-specific security data in these pipelines. 
This comparison of internal and external outcomes would also speak to a labor question of whether it could make sense for a firm to \textit{hire} the top external security researchers into a firm. Having shown that the bug bounty program was effective in attracting top researchers and high-value bugs, we have provided a natural launching point to study how to maximize the effectiveness of a third-party program relative to internal information sources. 

% \textbf{Retention of researchers.}
% While our analysis of veteran and new researchers was able to give some insight into whether veteran researchers maintained similar levels of outputs over time, we were not able to fully answer all questions around retention, including the longevity of researchers in the program. Ideally, it would be interesting to compare the average longevity of researchers in the program before and after the reward change; however, this requires data for a longer period of time after the reward change. Such analysis of retention could lead to valuable operational insights in a bug bounty program, especially if a program required significant amounts of training or onboarding of researchers.

\section{Rewards table snapshots}

\begin{figure*}[!ht]
    \centering
    \begin{tabular}{c}
         Before reward increase (May, 2024) \\
         \includegraphics[width=0.45\linewidth]{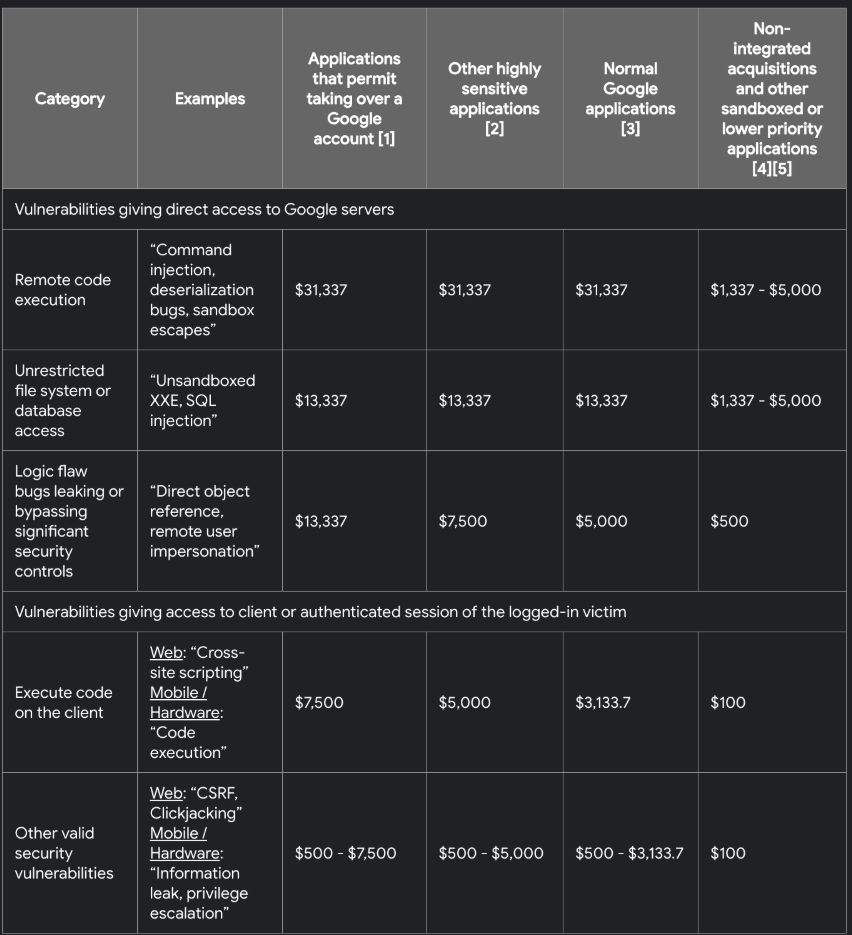} \\ After reward increase (July 2024) \\\includegraphics[width=0.45\linewidth]{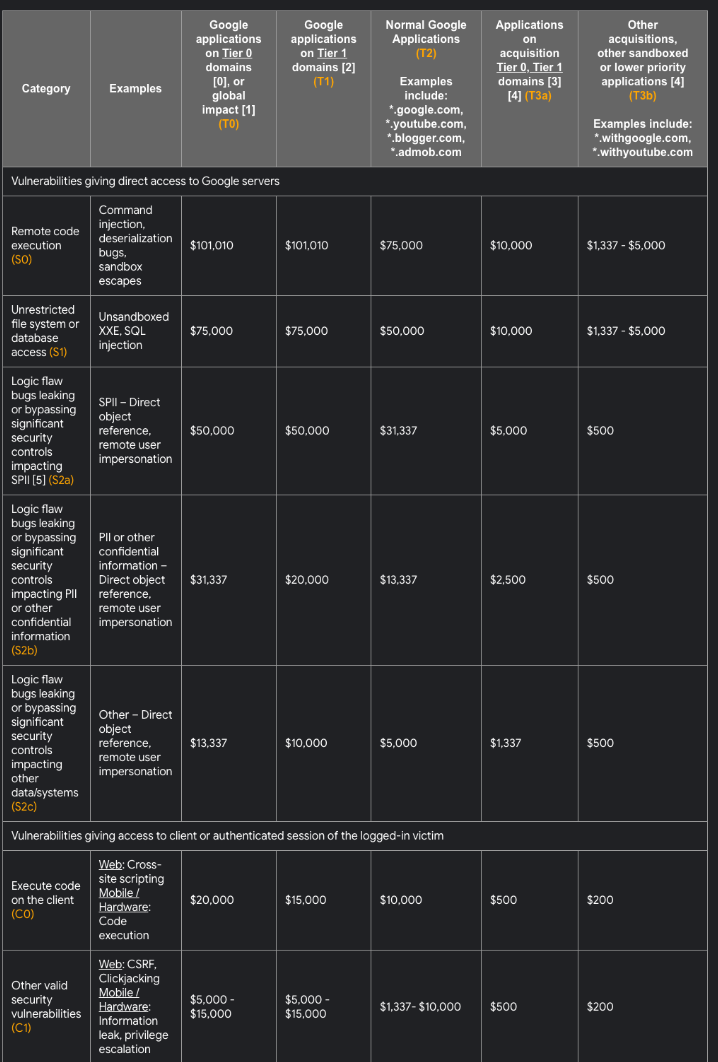}\\
    \end{tabular}
    \caption{Reward tables from before and after the GAVRP reward increase in July, 2024 (accessed via Wayback Machine). \textit{(Link: \url{https://bughunters.google.com/about/rules/google-friends/6625378258649088/google-and-alphabet-vulnerability-reward-program-vrp-rules})} }
    \label{fig:reward_tables}
\end{figure*}

\begin{figure*}[!ht]
    \centering
    \includegraphics[width=0.6\linewidth]{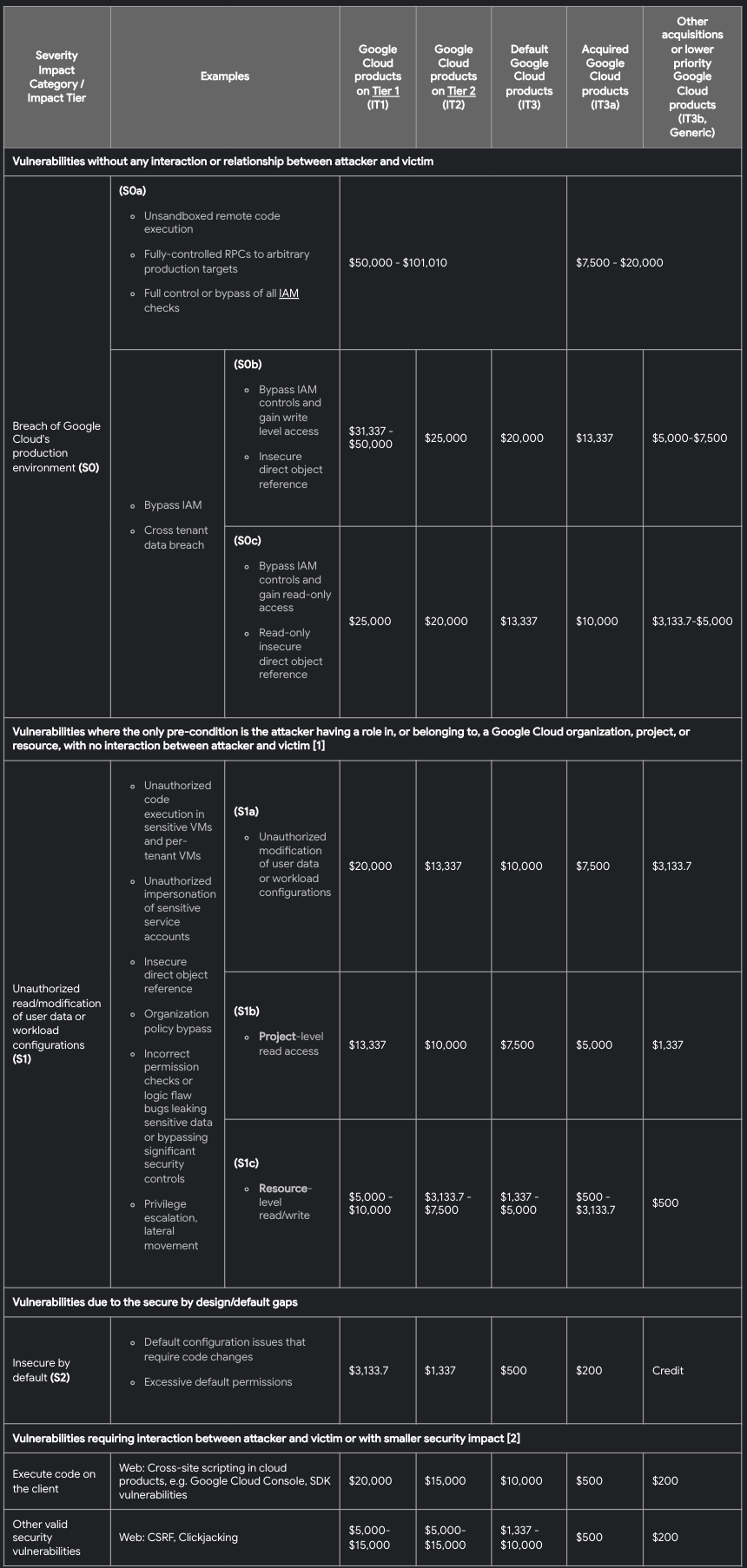} 
    \caption{Reward table for Cloud VRP. \textit{(Link: \url{https://bughunters.google.com/about/rules/google-friends/4849867320328192/cloud-vulnerability-reward-program-rules})}}
    \label{fig:reward_table_cloud}
\end{figure*}

\begin{figure*}[!ht]
    \centering
    \includegraphics[width=0.6\linewidth]{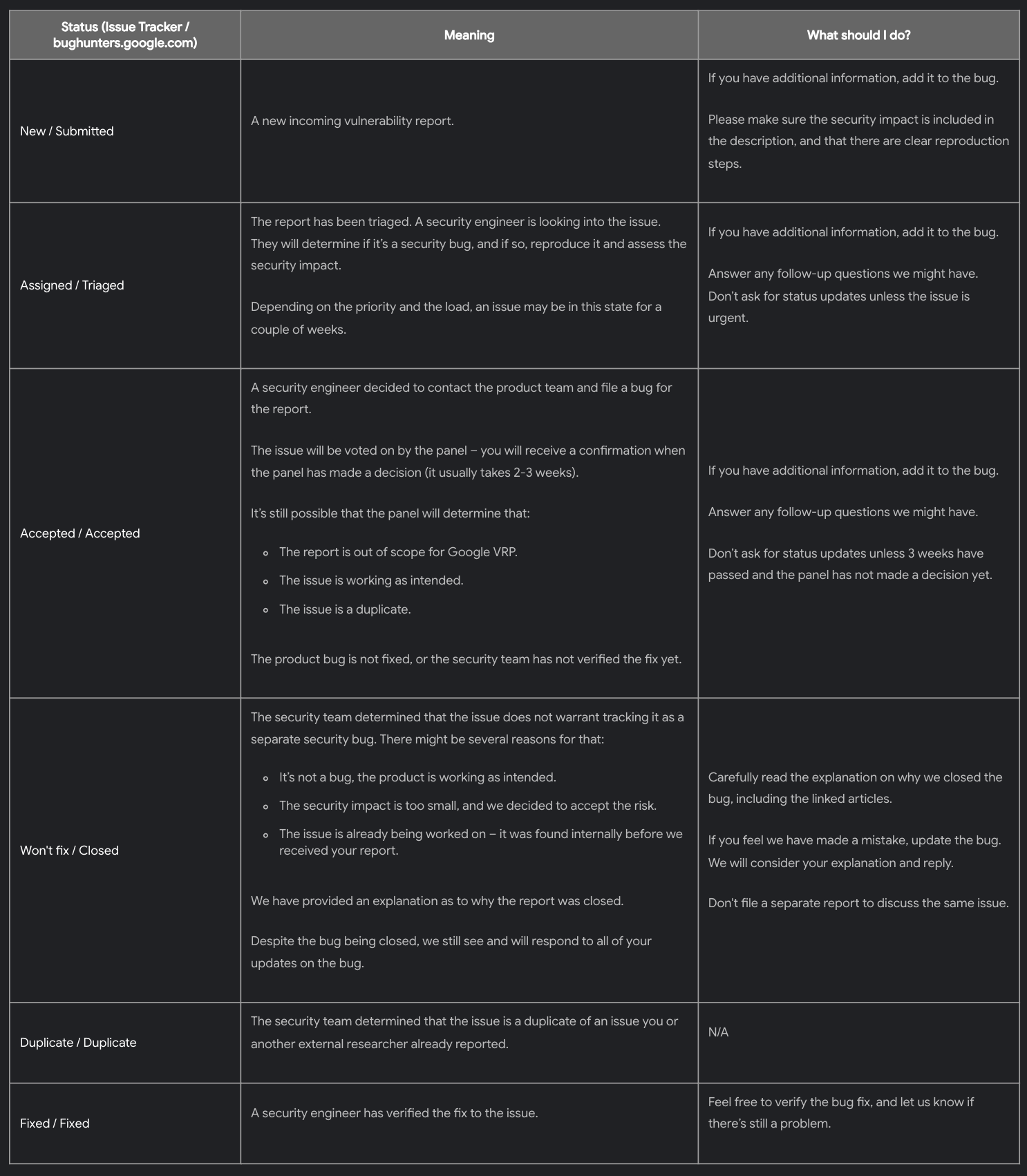} 
    \caption{Descriptions of various statuses that a submitted bug report can have. \textit{(Link: \url{https://bughunters.google.com/about/4925519884451840/frequently-asked-questions})}}
    \label{fig:report_status}
\end{figure*}

Figure \ref{fig:reward_tables} shows a snapshot of the GAVRP reward tables before and after the reward increase in July, 2024. The previous reward table was obtained via the Wayback Machine. Figure \ref{fig:reward_table_cloud} shows the reward table for the CVRP, which was created in October, 2024 branching from the original GAVRP, and offers reward amounts similar to the GAVRP rewards post-reward increase. Figure \ref{fig:report_status} shows various statuses that a submitted bug report can have.

The full rules and reward policies for the GAVRP, CVRP, OSSVRP, and AVRP programs can be found at \url{https://bughunters.google.com/about/rules/6744710187712512/about-this-section}.

\section{Removal of bugs from grants and events}
The VRPs occasionally run grants and special in-person events, often targeted at top researchers. To analyze the effects of the July 2024 reward change on the full population of researchers, we take a conservative approach of removing all bugs from the dataset that are associated with grants and in-person events. This would likely depress any observed treatment effects, as several grants and events were run after the reward change occurred that may have diverted researcher attention from their default program efforts. Still, we find effects in spite of the conservative approach.
% Still, we find effects in spite of the conservative approach of removing all bugs associated with grants and events.
% Notably, a large event and a grant, both for top researchers, took place in October and December 2024. Figure \ref{fig:all_timeseries_grants} in the Appendix shows the counts of all bugs with grants and events included. Still, we find effects in spite of the conservative approach of removing all bugs associated with grants and events.

\section{Additional results on quantity changes and distribution shift}

This section gives additional results on quantity changes and distribution shift across bug types.

\subsection{Observed elasticity for all bugs in treated and untreated programs}

Having observed that the mean bug count per month increased after the reward change, we compare this to the realized changes in average reward per bug and calculate an elasticity. Table \ref{tab:elasticities_all} shows the percent changes in quantity and realized reward, and the resulting estimated elasticity. For the treated programs, we observe an elasticity of $ \hat{\eta} = 0.206$, which indicates that a 100\% increase in paid rewards resulted in a roughly 20\% increase in the rate of bugs submitted per month. This fairly low elasticity indicates that a high reward increase is required to change the rate of bugs found per month overall, perhaps pointing to a barrier to entry for researchers, or a difficulty in capturing overall research effort. While the overall market might appear fairly inelastic, we show in Section \ref{sec:high-value} that the elasticity is significantly higher for high-value bugs. 

% We also compute an elasticity for the untreated program, but the confidence interval is extremely wide due to lack of statistical significance in both the change in $\bar{Y}$ and the change in $\bar{R}$.

\begin{table*}[!ht]
    \centering
    \begin{tabular}{c|c|c|c|c|c|c|c}
    \toprule
         Dataset & $\bar{Y}_0$ & $\bar{Y}_1$ & $\%\Delta \bar{Y}$ & $\bar{R}_0$ & $\bar{R}_1$ & $\%\Delta \bar{R}$ & $\hat{\eta}$ \\
         \midrule
         Treated & $35.38$ & $48.33$ & $36.57\%$ & $\$1,507.10$ &  $\$4,174.79$ & $177.00\%$ & $0.206$ \tiny{[CI: $(0.088, 0.451)$]} \\
         Untreated & $14.88$ & $16.5$ & $10.82\%$ & $\$1,783.44$ &  $\$3,142.24$ & $76.18\%$ & $0.142$ \tiny{[CI: $(-0.698, 1.803)$]} \\
    \bottomrule
    \end{tabular}
    \caption{Elasticity estimates for all bugs received in the treated and untreated programs.
    % Cross firm substitution? compare to amazon introduction in 2020
    % \textcolor{red}{write in text about realized rewards vs. table maximums; report category-weighted reward change}
    }
    \label{tab:elasticities_all}
\end{table*}

\subsection{Multiple hypothesis testing for distribution shift}\label{app:multiple_hypothesis_test}
We further dig into shifts towards or away from specific categories. For each type category $x$, let $p_x^0, p_x^1$ represent the probability before and after the reward change, respectively, that a bug belongs to category $x$. We run a two-sided t-test for the null hypothesis that $p^{0}_x = p^{1}_x$, generating a p-value and $95\%$ confidence interval for the difference in means, $\Delta \bar{p}_x = \bar{p}^1_x - \bar{p}^0_x$. We also compute a percent change,  $\% \Delta \bar{p}_x = \frac{\bar{p}^1_x - \bar{p}^0_x}{\bar{p}^0_x}$. We treat each type variable as a multiple hypothesis testing problem over $|\mathcal{X}|$ hypothesis tests, and apply the Holm-Bonferroni method \citep{holm1979simple} per type variable to control family-wise error rate at level $\alpha=0.05$. Table \ref{tab:types_t-test} shows increases in the proportions of high-value bugs for all type variables, with an over 600\% increase in the proportion of Tier 0 bugs, and a 176\% increase in High merit bugs. These results also suggest substitution of effort away from Medium Severity/Tier 2 bugs, but not from Low Severity/Tier 3 bugs.

\begingroup
\setlength{\tabcolsep}{4pt}
\begin{table}[!h]
    \centering
    \begin{tabular}[t]{c|c|c}
    \multicolumn{3}{c}{Tier} \\
    \toprule
    $x$ & $\Delta \bar{p}_x$ \footnotesize{($\% \Delta \bar{p}_x$)}& p-value \\
        \midrule
         \footnotesize{Tier 0} &  \textcolor{darkgreen}{\textbf{0.059 \footnotesize{(628\%)}}}  & \textbf{3.36e-7} \\
         \footnotesize{Tier 1} &  \textcolor{darkgreen}{\textbf{0.052  \footnotesize{(104\%)}}} & \textbf{0.0028} \\
         \footnotesize{Tier 2} & \textcolor{red}{\textbf{-0.152  \footnotesize{(-52\%)}}} & \textbf{4.38e-7}  \\
         \footnotesize{Tier 3} & \textcolor{darkgreen}{0.040 \footnotesize{(24\%)}} & 0.129  \\
        \bottomrule
    \end{tabular} \,
    \begin{tabular}[t]{c|c|c} 
    \multicolumn{3}{c}{Severity} \\
    \toprule
    $x$ & $\Delta \bar{p}_x$ \footnotesize{($\% \Delta \bar{p}_x$)} & p-value \\
        \midrule
         \footnotesize{Crit} &  \textcolor{darkgreen}{0.002 \footnotesize{(9\%)}}  & 0.836 \\
         \footnotesize{High} &  \textcolor{darkgreen}{0.037  \footnotesize{(14\%)}} & 0.233 \\
         \footnotesize{Med} & \textcolor{red}{-0.072  \footnotesize{(-11\%)}} & 0.028  \\
         \footnotesize{Low} & \textcolor{darkgreen}{\textbf{0.032 \footnotesize{(150\%)}}} & \textbf{0.008}  \\
        \bottomrule
    \end{tabular} \,
    \begin{tabular}[t]{c|c|c} 
    \multicolumn{3}{c}{Merit} \\
    \toprule
    $x$ & $\Delta \bar{p}_x$ \footnotesize{($\% \Delta \bar{p}_x$)} & p-value \\
        \midrule
         \footnotesize{High} &  \textcolor{darkgreen}{\textbf{0.074 \footnotesize{(176\%)}}}  & \textbf{1.89e-5} \\
        \bottomrule
    \end{tabular}
    \caption{Changes in category proportions for each bug type variable. Rows identified as significant via the Holm-Bonferroni method at level $\alpha = 0.05$ per type variable are in bold.}
    \label{tab:types_t-test}
\end{table}
\endgroup

\subsection{Observed quantity changes for all bug types}

\begin{figure*}[!h]
    \centering
    \begin{tabular}{ccc}
    \small{Tier (treated program)} & \small{Severity (treated program)}  & \small{Merit (treated program)} \\
    \begin{minipage}{0.3\textwidth}
    \centering
    \begin{tikzpicture}
      \node (img)  {\includegraphics[scale=0.3]{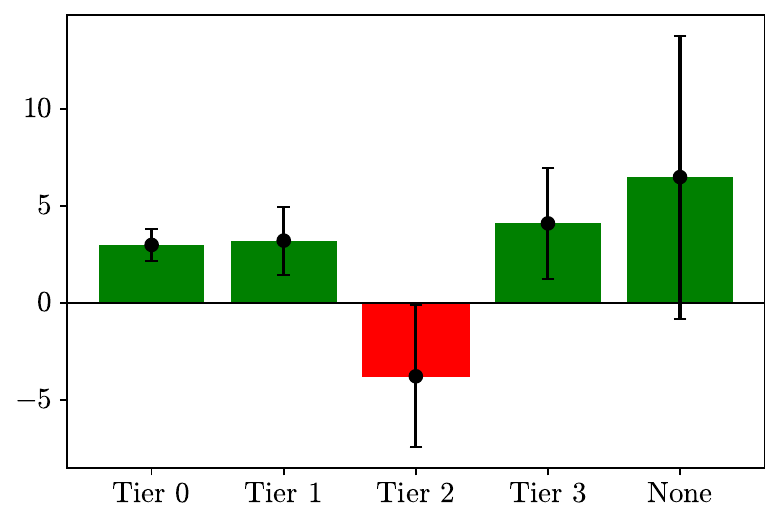}};
    %   \node[left=of img, node distance=0cm, rotate=0, anchor=center,yshift=0cm,xshift=0.9cm] {$\hat{\tau}^{\text{DiD}}_{x}$};
     \end{tikzpicture}
    \end{minipage} & 
    \begin{minipage}{0.3\textwidth}
    \centering
    \begin{tikzpicture}
      \node (img)  {\includegraphics[scale=0.3]{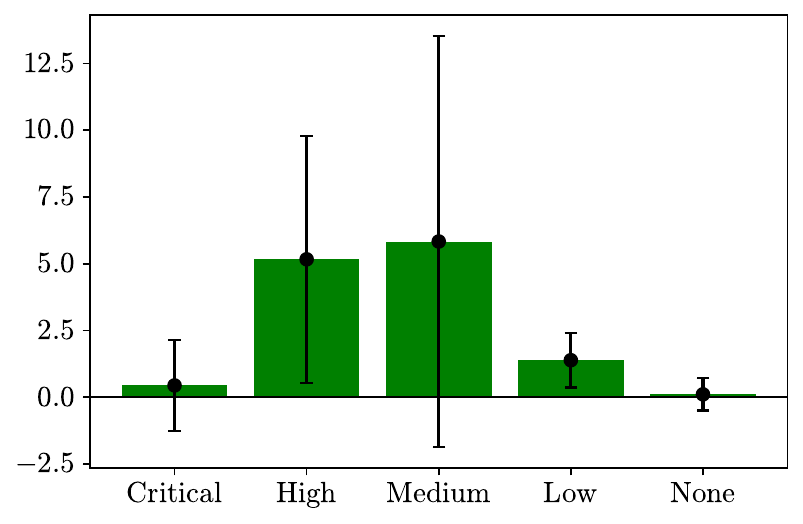}};
    \end{tikzpicture}
    \end{minipage} &
    \begin{minipage}{0.3\textwidth}
    \centering
    \begin{tikzpicture}
      \node (img)  {\includegraphics[scale=0.3]{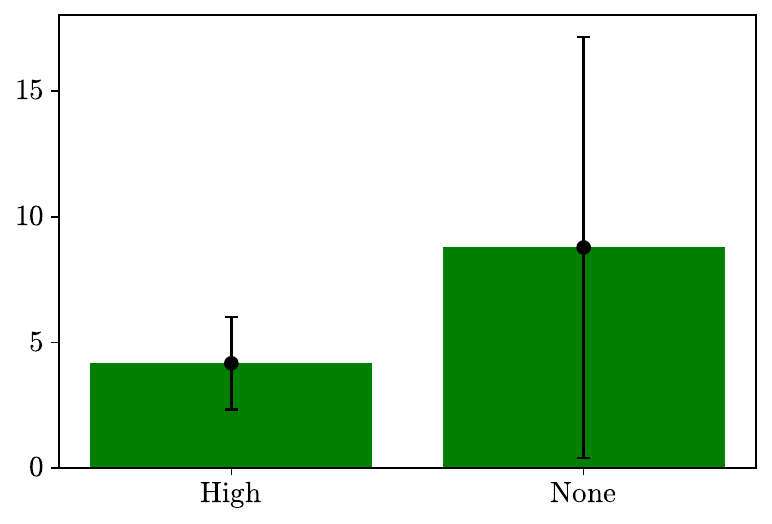}};
     \end{tikzpicture}
    \end{minipage} 
     \\
    \small{Tier (untreated programs)} & \small{Severity (untreated programs)}  & \small{Merit (untreated programs)} \\
    \begin{minipage}{0.3\textwidth}
    \centering
    \begin{tikzpicture}
      \node (img)  {\includegraphics[scale=0.3]{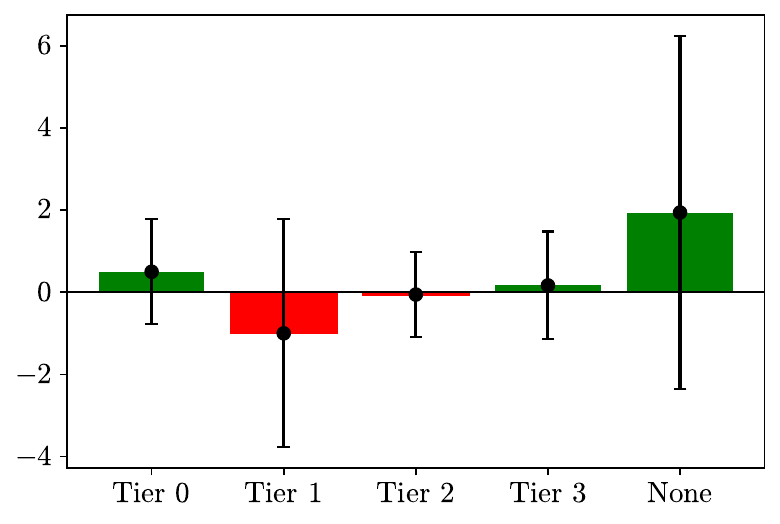}};
    %   \node[left=of img, node distance=0cm, rotate=0, anchor=center,yshift=0cm,xshift=0.9cm] {$\hat{\tau}^{\text{DiD}}_{x}$};
     \end{tikzpicture}
    \end{minipage} & 
    \begin{minipage}{0.3\textwidth}
    \centering
    \begin{tikzpicture}
      \node (img)  {\includegraphics[scale=0.3]{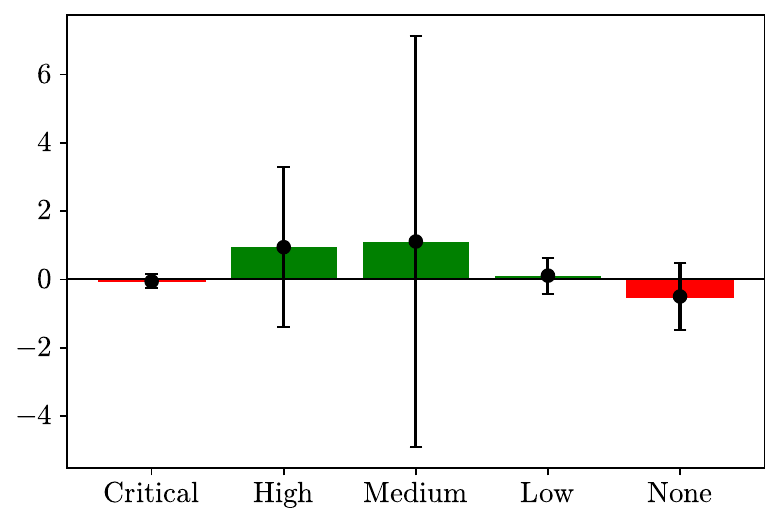}};
    \end{tikzpicture}
    \end{minipage} &
    \begin{minipage}{0.3\textwidth}
    \centering
    \begin{tikzpicture}
      \node (img)  {\includegraphics[scale=0.3]{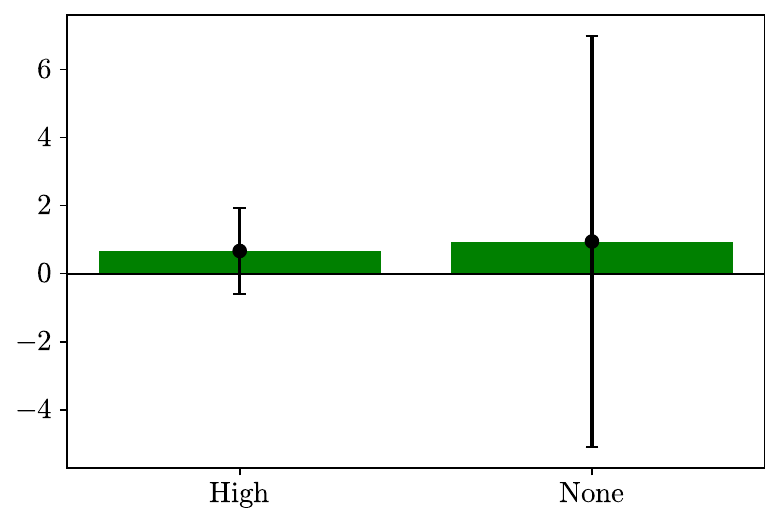}};
     \end{tikzpicture}
    \end{minipage} \\
    \end{tabular}
    \caption{Difference in mean bugs per month $\Delta \bar{Y}_x$ for all tier, severity, and merit types in the treated and untreated programs.
    }
    \label{fig:diff_all_types}
\end{figure*}

In addition to the high-value bug types analyzed in Section \ref{sec:high-value}, Figure \ref{fig:diff_all_types} also shows the change in mean bugs per month for all other types. This differs from the distributional analysis in Section \ref{sec:distribution} in that $\Delta\bar{Y}_x$ measures total bug counts per month, rather than relative quantities measured by $\Delta \bar{p}_x$.

% \subsubsection{\textcolor{blue}{Multiple hypothesis testing procedure}}
% \textcolor{blue}{For each type variable $X$ with domain $\mathcal{X}$, we test for the change in mean bug counts for month $\Delta \bar{Y}_x$ for each type category $x \in \mathcal{X}$. The difference between this and the distributional analysis is that we now consider raw counts rather than relative frequencies. This yields $|\mathcal{X}|$ tests and p-values. We provide all p-values for each individual test, and employ the Holm-Bonferroni method to control family-wise error rate at level $\alpha = 0.05$ for each type variable.}

% \section{Grants and events}

% % We provide additional figures to illustrate robustness checks with alternative data configurations.

% Figure \ref{fig:all_timeseries_grants} shows all bugs per month received from 2023-2024 in the treated program, including bugs from grants and events.

% \begin{figure}[!ht]
%     \centering
%     \begin{minipage}{0.4\textwidth}
%     \centering
%     \begin{tikzpicture}
%       \node (img)  {\includegraphics[scale=0.4]{figures/all_timeseries_gc_grants.pdf}};
%       \node[above=of img, node distance=0cm, rotate=0, anchor=center,yshift=-0.9cm,xshift=0cm] {\footnotesize{Treated program (with grants and events)}};
%      \end{tikzpicture}
%      \end{minipage}
%     \caption{All bugs per month received from 2023-2024 in the treated program, including bugs from grants and events.}
%     \label{fig:all_timeseries_grants}
% \end{figure}

\section{Additional results on researcher participation}

This section gives additional results surrounding researcher participation, including the participation of veteran vs. new researchers.

\subsection{Number of new successful researchers}

\begin{figure}[!ht]
    \centering
    \begin{minipage}{0.45\textwidth}
    \centering
    \begin{tikzpicture}
      \node (img)  {\includegraphics[scale=0.35]{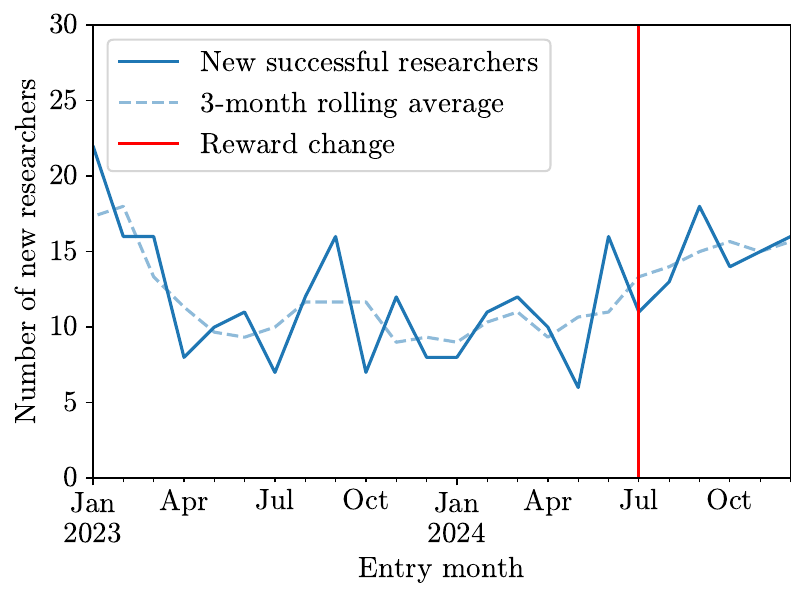}};
      \node[above=of img, node distance=0cm, rotate=0, anchor=center,yshift=-0.9cm,xshift=0cm] {\scriptsize{Number of new successful researchers}};
     \end{tikzpicture}
    \end{minipage} 
    \caption{Number of new researchers each month who submit a report for the first time to the treated program, and successfully find a product bug in their first six months of participation.}
    \label{fig:new_successful}
\end{figure}

The vast majority of new researchers never successfully find a product bug. Filtering for ``successful'' researchers, Figure \ref{fig:new_successful} shows the number of new researchers who find at least one product bug in their first six months of submitting bug reports.

\subsection{Researcher substitution across programs}\label{app:substitution}
\begin{figure}[!ht]
    \centering
    \begin{tabular}[t]{cc}
    \begin{minipage}{0.45\textwidth}
    % \centering
    \begin{tikzpicture}
      \node (img)  {\includegraphics[scale=0.34]{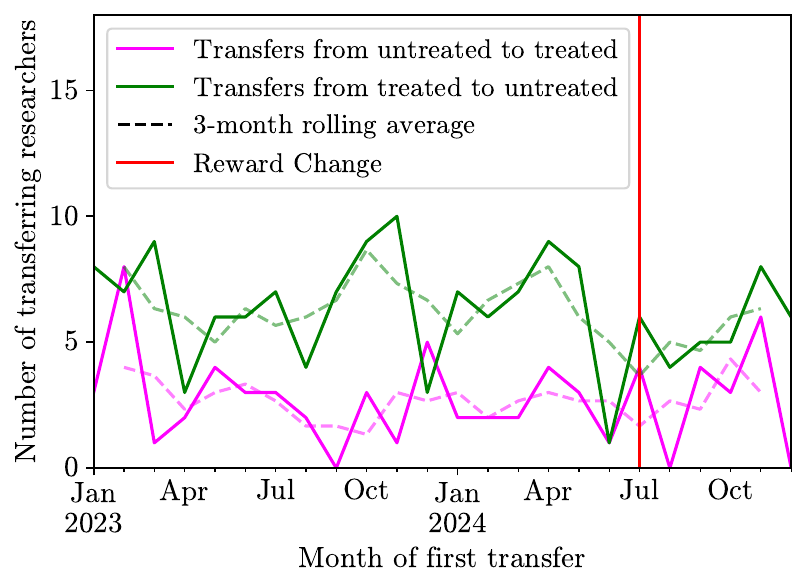}};
      \node[above=of img, node distance=0cm, rotate=0, anchor=center,yshift=-0.9cm,xshift=0cm] {\scriptsize{Number of transferring researchers}};
     \end{tikzpicture}
    \end{minipage} & 
    \begin{minipage}{0.45\textwidth}
    % \centering
    \begin{tikzpicture}
      \node (img)  {\includegraphics[scale=0.34]{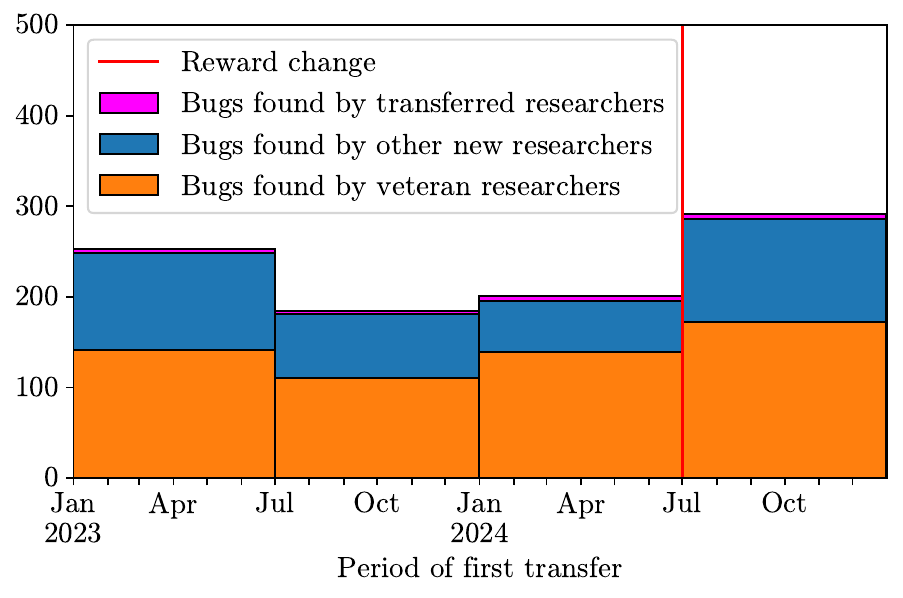}};
      \node[above=of img, node distance=0cm, rotate=0, anchor=center,yshift=-0.95cm,xshift=0cm] {\scriptsize{Shares of bugs found by transferred researchers}};
     \end{tikzpicture}
    \end{minipage} 
    \end{tabular}
    \caption{\textit{(Left)} Number of researchers each month who transfer and submit a report to a different program for the first time. \textit{(Right)} Shares of bugs found by transferred researchers in the treated program. }
    \label{fig:transfer}
\end{figure}

A natural question is how many of the new researchers are really fully ``new,'' and how many may have transferred from the untreated programs. In general, we find that only a handful of researchers transferred between programs at all, and Figure \ref{fig:transfer} \textit{(left)} shows that the rates of transfer from the untreated program to the treated program and vice versa did not seem affected by the reward change. Furthermore, the share of bugs found by transferred researchers was negligibly small (Figure \ref{fig:transfer} (\textit{right})). This matches domain expert intuition that substitution across these programs is difficult, likely due to differences in the software covered by the different programs. 
% This also suggests that the new researchers in the treated program could be coming from working on similar software domains, perhaps in other external programs.

\section{Robustness Check: Removal of Cloud bugs}\label{app:nocloud}

As a robustness check to analyze a setting without the introduction of the CVRP in October, 2024, we run the same analyses for bugs in the GAVRP that excludes all bugs related to the Cloud product both before and after the introduction of the CVRP. There are several important drawbacks to this analysis that limit the conclusions that can be drawn.
\begin{itemize}
    \item Cloud-related bugs made up roughly 40\% of bugs collected by the GAVRP prior to split of the CVRP. Thus, data for non-Cloud bugs in the GAVRP removes a significant portion of key researchers and bugs of practical interest.
    \item It is difficult to disentangle whether any observed increase in Cloud bugs is due to a delayed impact from the reward change in July, or more publicity for higher rewards, as the CVRP employs similar reward amounts to the new GAVRP reward amounts after the reward increase. There is no counterfactual for an announcement for a CVRP without the reward increase.
    \item There was a large grant offered to a small group of top GAVRP researchers in December, 2024, which drew top researcher attention away from regular GAVRP activity in the period after the CVRP launch. This grant was not offered in the CVRP.
\end{itemize}

For the GAVRP with Cloud bugs removed, we still observe increases in high-value bugs after the reward change, and we do not observe increases for all bugs. These findings roughly match the findings from the full dataset without Cloud bugs removed, except with less of an increase in the counts of all bugs per month. The lack of increase in all bugs per month might be due in part to the large grant in the GAVRP at the end of 2024. We provide full detailed analysis below.

% Given the large grant in the GAVRP at the end of 2024, we cannot definitively attribute this discrepancy in total bug counts to the Cloud announcement. 

% There is also currently not enough data after October, 2024 to run a high-powered DiD design to confirm an effect of the announcement of the CVRP. However, such a test may be possible in the future with more data from 2025.

\subsection{Dataset summary} 

Our dataset the \textit{treated program without Cloud bugs} consists of all de-duplicated product bugs classified as vulnerabilities received by the GAVRP between January, 2023 and December, 2024, and excludes all bugs assigned by Google-internal security engineers to a product within the Google Cloud product area. As before, all bugs related to grants and events are removed. This yields a total of 500 bugs from 266 distinct researchers.

\subsection{Results}

\begin{figure}[!ht]
    \centering
    \begin{minipage}{0.4\textwidth}
    \centering
    \begin{tikzpicture}
      \node (img)  {\includegraphics[scale=0.4]{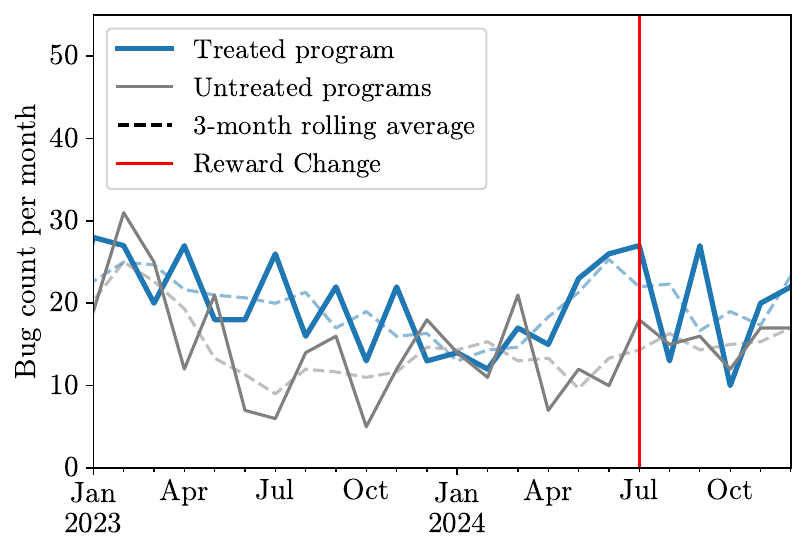}};
      \node[above=of img, node distance=0cm, rotate=0, anchor=center,yshift=-0.9cm,xshift=0cm] {\footnotesize{Treated program (no Cloud)}};
     \end{tikzpicture}
     \end{minipage}
    \caption{All bugs per month received in the treated program without Cloud bugs vs. the untreated programs.}
    \label{fig:all_timeseries_nocloud}
\end{figure}

\paragraph{All bugs.} Removing all Cloud bugs, we no longer observe an increase in the overall bug counts (Figure \ref{fig:all_timeseries_nocloud}). Plausible explanations for this discrepancy include an effect of the Cloud announcement, or an effect of the GAVRP grant. 

\begin{figure*}[!ht]
    \centering
    \begin{tabular}{ccc}
    \scriptsize{Tier 0 (no Cloud)} & \scriptsize{Severity $\geq$ High (no Cloud)} & \scriptsize{High Merit (no Cloud)}  \\
    \begin{minipage}{0.3\textwidth}
    \centering
    \begin{tikzpicture}
      \node (img)  {\includegraphics[scale=0.3]{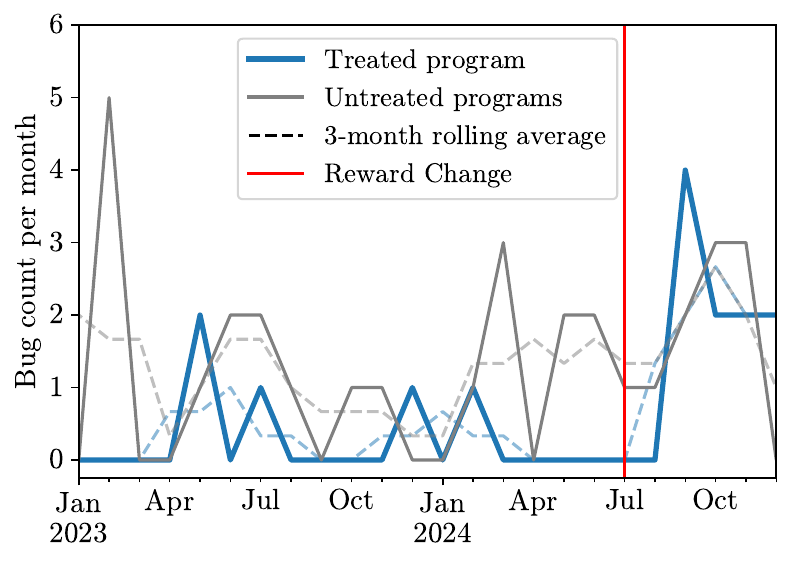}};
     \end{tikzpicture}
    \end{minipage} & 
    \begin{minipage}{0.3\textwidth}
    \centering
    \begin{tikzpicture}
      \node (img)  {\includegraphics[scale=0.3]{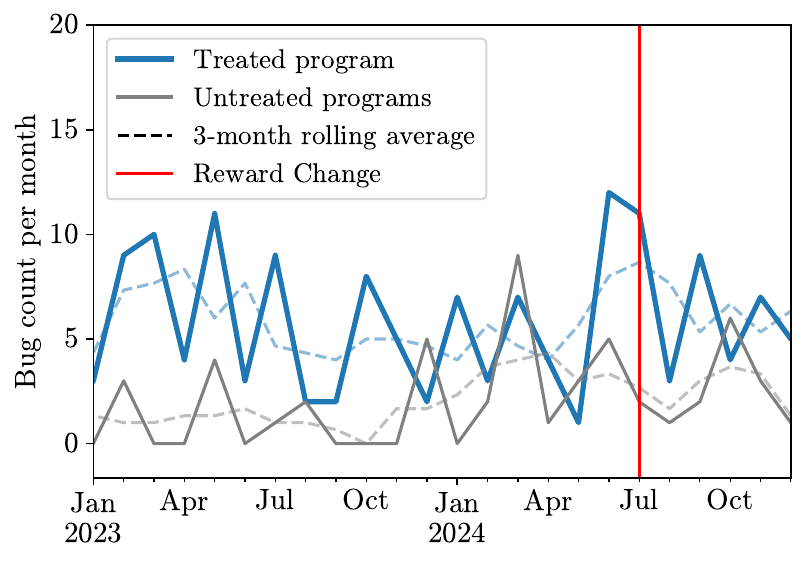}};
    \end{tikzpicture}
    \end{minipage} &
    \begin{minipage}{0.3\textwidth}
    \centering
    \begin{tikzpicture}
      \node (img)  {\includegraphics[scale=0.3]{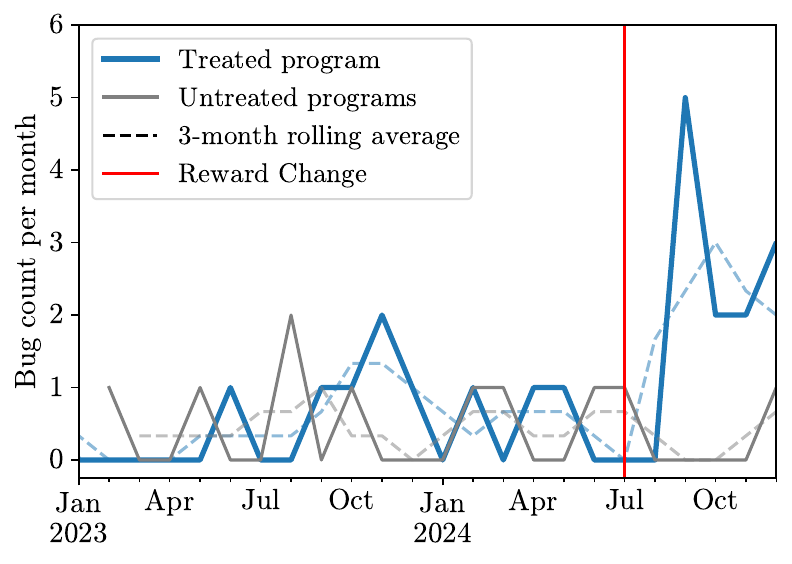}};
     \end{tikzpicture}
    \end{minipage} 
    \end{tabular}
    \caption{Monthly bug counts of high-value types for tier, severity, and merit received in the treated program without Cloud bugs before and after the reward change in July, 2024. 
    }
    \label{fig:type_timeseries_nocloud}
\end{figure*}

\begin{figure*}[!ht]
    \centering
    \begin{tabular}{ccc}
    \scriptsize{Tier (no Cloud)} & \scriptsize{Severity (no Cloud)}  & \scriptsize{Merit (no Cloud)} \\
    \begin{minipage}{0.35\textwidth}
    \centering
    \begin{tikzpicture}
      \node (img)  {\includegraphics[scale=0.35]{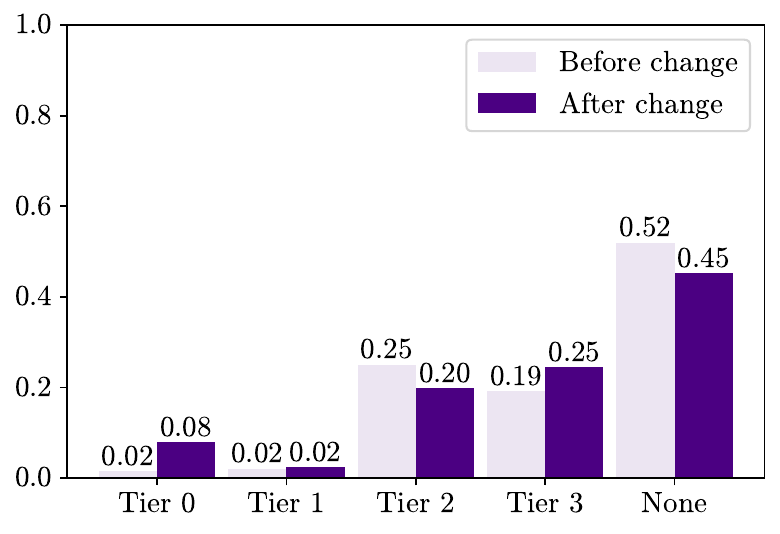}};
     \end{tikzpicture}
    \end{minipage} & 
    \begin{minipage}{0.35\textwidth}
    \centering
    \begin{tikzpicture}
      \node (img)  {\includegraphics[scale=0.35]{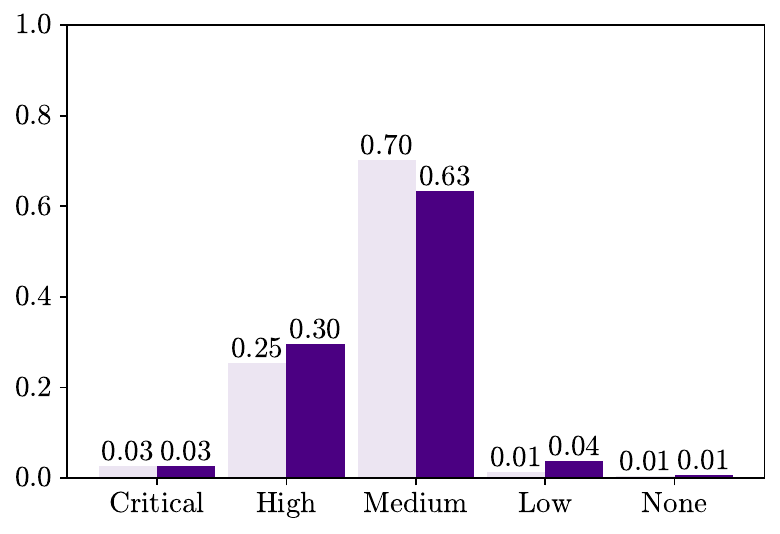}};
    \end{tikzpicture}
    \end{minipage} &
    \begin{minipage}{0.2\textwidth}
    \centering
    \begin{tikzpicture}
      \node (img)  {\includegraphics[scale=0.28]{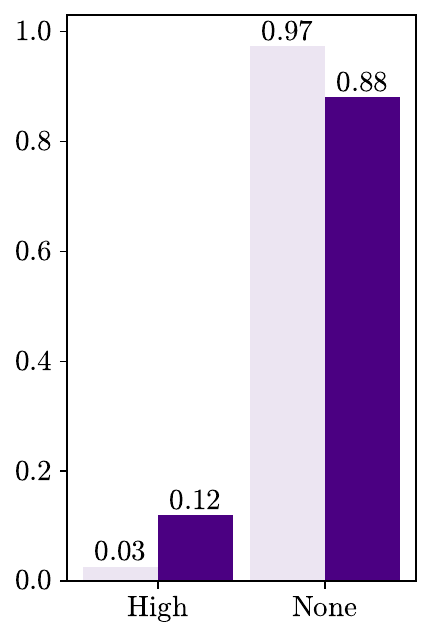}};
     \end{tikzpicture}
    \end{minipage} 
    \end{tabular}
    \caption{Distributions of tier, severity, and merit for all bugs received in the treated program without Cloud bugs before and after the reward change in July, 2024. 
    % Results of mean estimation and t-test: $\bar{p}_0^{\text{Tier}} = 0.011$, $\bar{p}_1^{\text{Tier}} = 0.066$, $\Delta \bar{p}^{\text{Tier}} = 0.055$, $95\%$ CI: $(0.033, 0.078)$, p-value: $2.14e-6$ (reject null). 
    % \textcolor{red}{proportions instead of counts}
    }
    \label{fig:type_hists_nocloud}
\end{figure*}

\begin{figure*}[!ht]
    \centering
    \begin{tabular}{ccc}
    \scriptsize{Tier (treated program, no Cloud)} & \scriptsize{Severity (treated program, no Cloud)}  & \scriptsize{Merit (treated program, no Cloud)} \\
    \begin{minipage}{0.3\textwidth}
    \centering
    \begin{tikzpicture}
      \node (img)  {\includegraphics[scale=0.3]{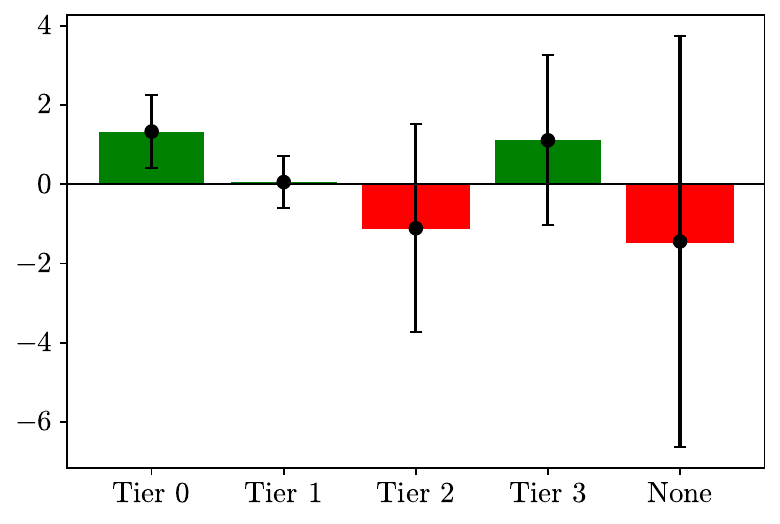}};
    %   \node[left=of img, node distance=0cm, rotate=0, anchor=center,yshift=0cm,xshift=0.9cm] {$\hat{\tau}^{\text{DiD}}_{x}$};
     \end{tikzpicture}
    \end{minipage} & 
    \begin{minipage}{0.3\textwidth}
    \centering
    \begin{tikzpicture}
      \node (img)  {\includegraphics[scale=0.3]{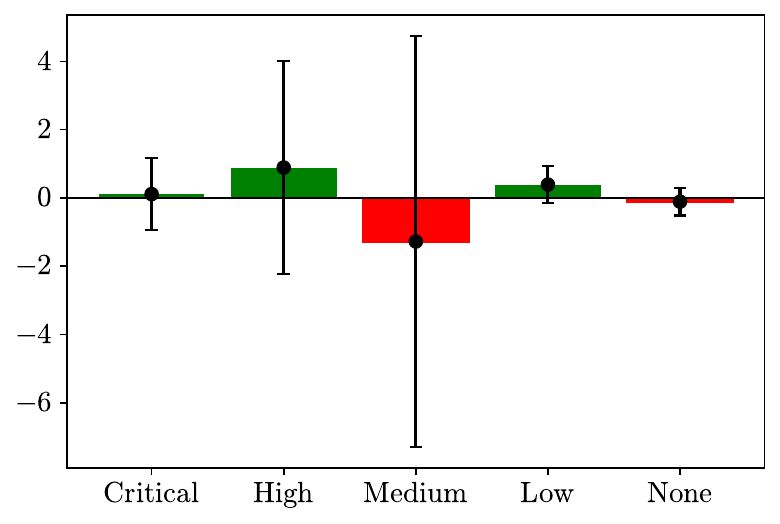}};
    \end{tikzpicture}
    \end{minipage} &
    \begin{minipage}{0.3\textwidth}
    \centering
    \begin{tikzpicture}
      \node (img)  {\includegraphics[scale=0.3]{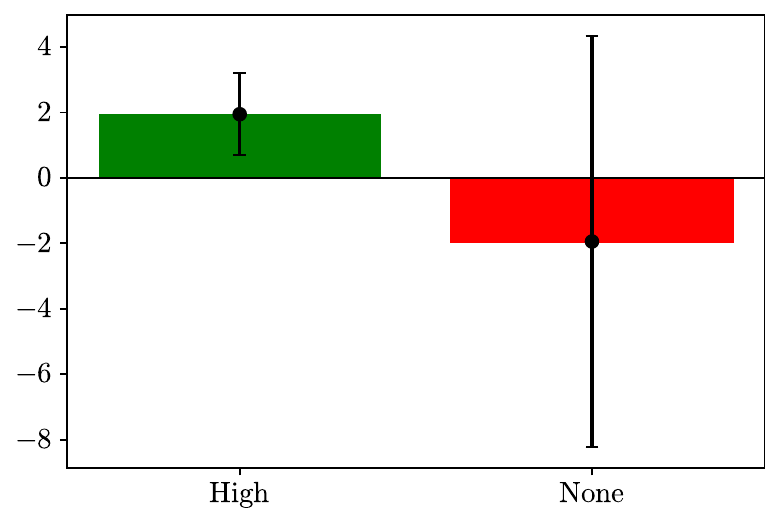}};
     \end{tikzpicture}
    \end{minipage} 
     \\
    \end{tabular}
    \caption{Difference in observed mean bugs per month $\Delta \bar{Y}_x$ for all tier, severity, and merit types in the treated program without Cloud bugs.
    }
    \label{fig:diff_all_types_nocloud}
\end{figure*}

\paragraph{High-value bugs.} As with the treated program as a whole, we observe significant increases the rate of bugs received for month for high-value bugs of high tier and merit (Figures \ref{fig:type_timeseries_nocloud}). We also observe a shift in the distribution of found bugs towards higher tiers, severities, and merit (Figure \ref{fig:type_hists_nocloud}). Figure \ref{fig:diff_all_types_nocloud} shows that there is a statistically significant increase in the observed mean bug count per month for Tier 0 and High Merit bugs. 

% \begin{figure*}[!ht]
%     \centering
%     \begin{tabular}{ccc}
%     Tier (no Cloud) & Severity (no Cloud)  & Merit (no Cloud) \\
%     \begin{minipage}{0.3\textwidth}
%     \begin{tikzpicture}
%       \node (img)  {\includegraphics[scale=0.3]{figures/did_tier_prior_3_nocloud.pdf}};
%     %   \node[left=of img, node distance=0cm, rotate=0, anchor=center,yshift=0cm,xshift=0.9cm] {$\hat{\tau}^{\text{DiD}}_{x}$};
%      \end{tikzpicture}
%     \end{minipage} & 
%     \begin{minipage}{0.3\textwidth}
%     \begin{tikzpicture}
%       \node (img)  {\includegraphics[scale=0.3]{figures/did_severity_prior_3_nocloud.pdf}};
%     \end{tikzpicture}
%     \end{minipage} &
%     \begin{minipage}{0.3\textwidth}
%     \begin{tikzpicture}
%       \node (img)  {\includegraphics[scale=0.3]{figures/did_merit_prior_3_nocloud.pdf}};
%      \end{tikzpicture}
%     \end{minipage} 
%     \end{tabular}
%     \caption{Difference in difference estimates $\hat{\tau}^{\text{DiD}}_{x}$ for all tier, severity, and merit types in the treated program without Cloud bugs.
%     }
%     \label{fig:did_all_types_nocloud}
% \end{figure*}

\newpage
\section{Extended effects into 2025}\label{app:2025}

We intentionally limited our analysis in the main paper to short-term effects following the reward change in July, 2024, as the months following enjoyed a uniquely quiet period in terms of external events. We continued to observe persistent increases in bug counts in 2025; however, it becomes increasingly hard to say with any certainty that these increases can be attributed to the reward change in July, 2024. As discussed in the Limitations section, several exogenous factors in 2025 include a rise in popularity of AI tools for bug hunting and writing bug reports \citep{aireports}, the introduction of new AI bug bounty programs and significant reward increases outside Google \citep{anthropic,openaiincrease,appleincrease,google_ai_bug_bounty}, and the addition of AI products as software targets to the Google VRPs. The analysis of the effects of these events remains an open question. We report bug counts through September, 2025 here as a reference, and briefly discuss possible implications.
%, but do not consider this for rigorous statistical inference.

\begin{figure*}[!ht]
    \centering
    \includegraphics[scale=0.4]{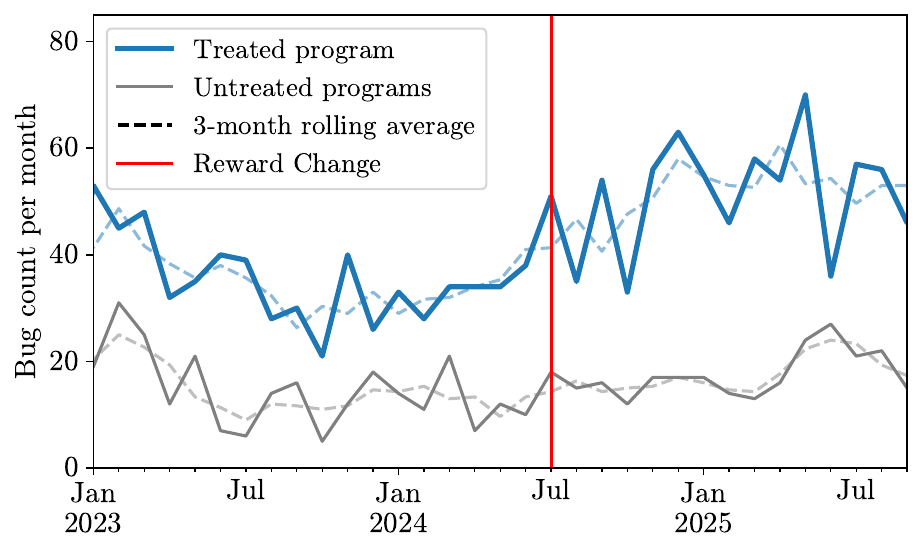}
    \caption{All bugs per month received in the treated vs. untreated programs through September, 2025.}
    \label{fig:all_timeseries_2025}
\end{figure*}

Figure \ref{fig:all_timeseries_2025} shows that the total number of bugs found in the treated program continued to grow into 2025. However, there was also an increase in bugs found in the untreated programs in 2025, due in part to the introduction of AI products as software targets into the AVRP in 2025. Thus, it is harder to think of the AVRP as a valid control in 2025. 

In terms of high-value types, Figure \ref{fig:type_timeseries_2025} shows that the number of Tier 0 bugs and high merit bugs seemed to taper off in later 2025. This could potentially indicate that the difficulty in finding these high-value bugs increased into 2025 after many were found in late 2024. In the untreated programs, there is also an increase in High Merit bugs in 2025. Most of these additional High Merit bugs are for AI products introduced as software targets into the AVRP in 2025, again reflecting the difficulty of using the AVRP as a control in 2025.

\begingroup
\setlength{\tabcolsep}{2pt}
\begin{figure*}[!ht]
    \centering
    \begin{tabular}{ccc}
    % \scriptsize{Tier 0} & \scriptsize{Severity $\geq$ High} & \scriptsize{High Merit}  \\
    \begin{minipage}{0.32\textwidth}
    \centering
    \begin{tikzpicture}
      \node (img)  {\includegraphics[scale=0.29]{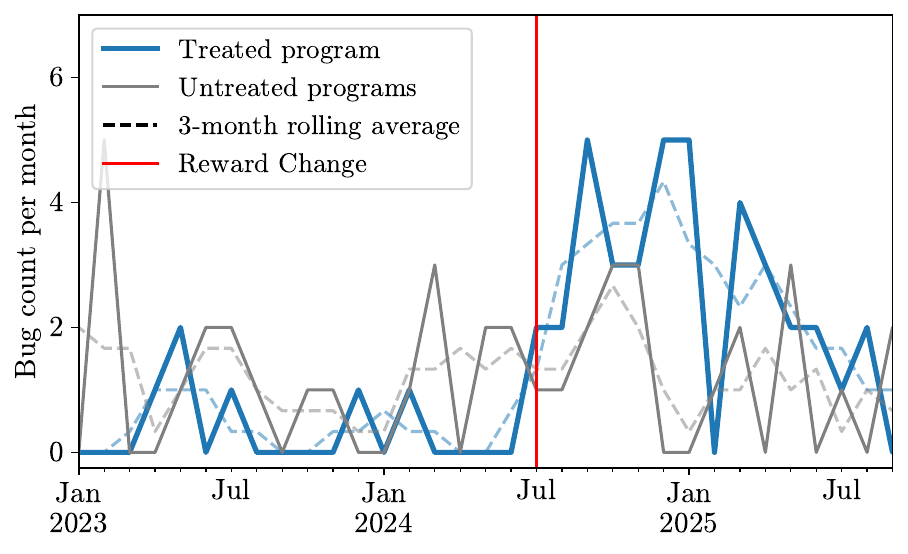}};
      \node[above=of img, node distance=0cm, rotate=0, anchor=center,yshift=-0.9cm,xshift=0cm, text width=\textwidth, align=center] {\scriptsize{Tier 0}};
     \end{tikzpicture}
    \end{minipage} & 
    \begin{minipage}{0.32\textwidth}
    \centering
    \begin{tikzpicture}
      \node (img)  {\includegraphics[scale=0.29]{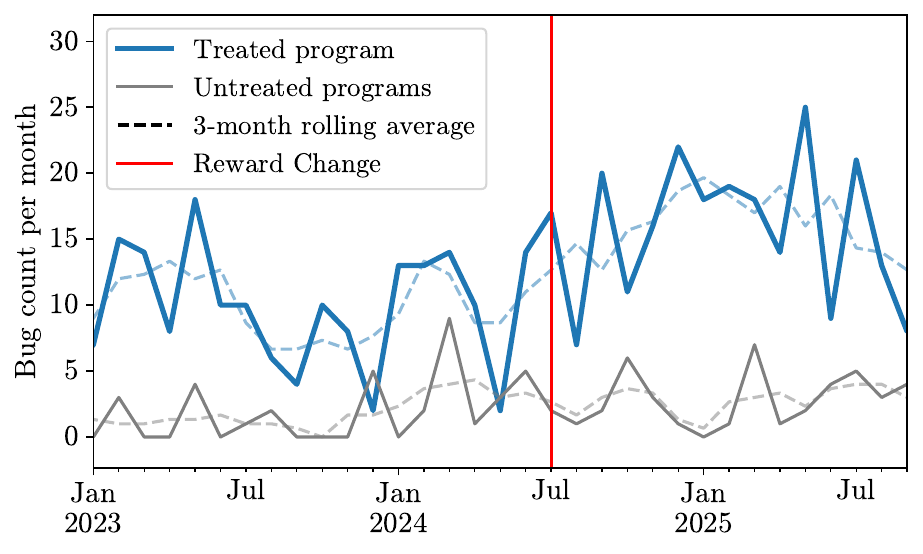}};
      \node[above=of img, node distance=0cm, rotate=0, anchor=center,yshift=-0.9cm,xshift=0cm, text width=\textwidth, align=center] {\scriptsize{Severity $\geq$ High}};
    \end{tikzpicture}
    \end{minipage} &
    \begin{minipage}{0.32\textwidth}
    \centering
    \begin{tikzpicture}
      \node (img)  {\includegraphics[scale=0.29]{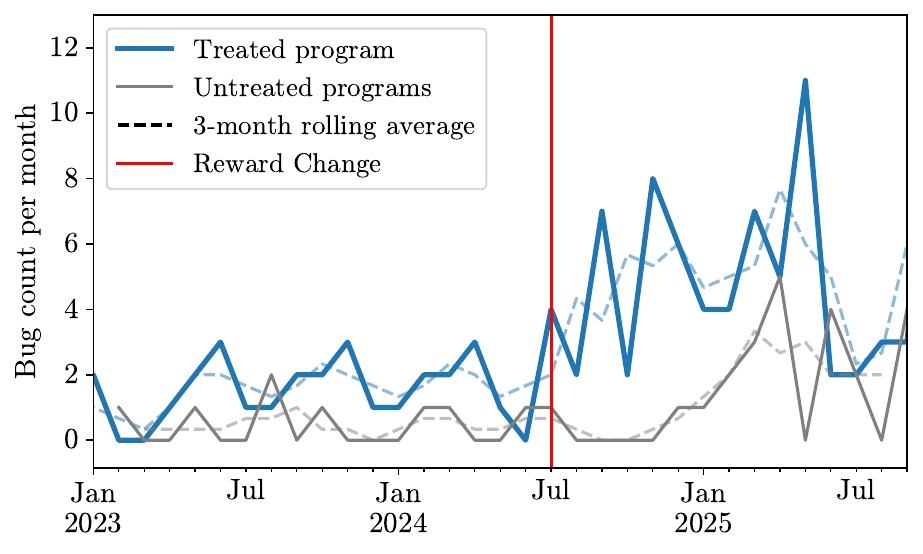}};
      \node[above=of img, node distance=0cm, rotate=0, anchor=center,yshift=-0.9cm,xshift=0cm, text width=\textwidth, align=center] {\scriptsize{High Merit}};
     \end{tikzpicture}
    \end{minipage} 
    \end{tabular}
    \caption{High-value bugs per month received in the treated vs. untreated programs through September, 2025.
    }
    \label{fig:type_timeseries_2025}
\end{figure*}
\endgroup

\begin{figure}[!ht]
    \centering
    \begin{tabular}{cc}
    \begin{minipage}{0.45\textwidth}
    \centering
    \begin{tikzpicture}
      \node (img)  {\includegraphics[scale=0.38]{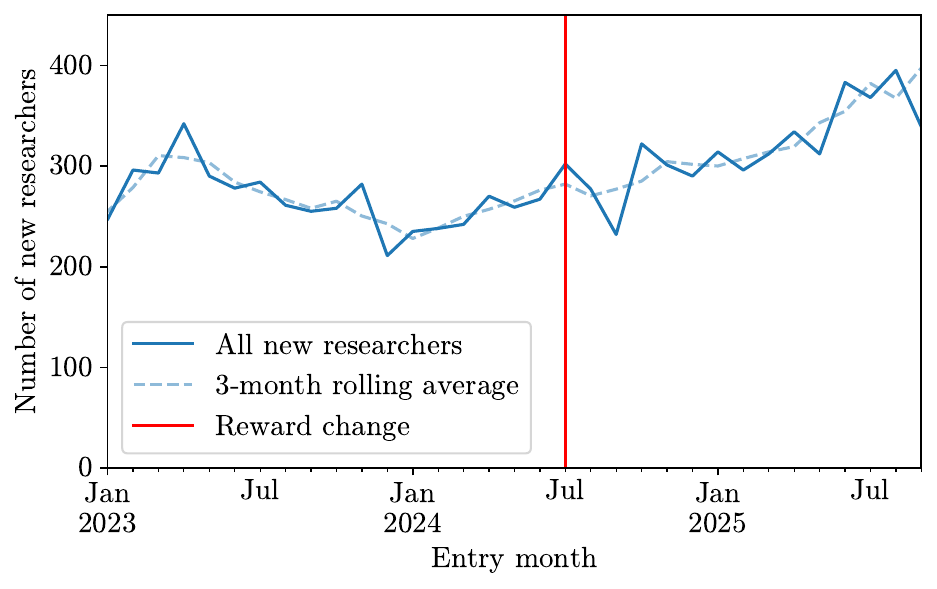}};
      \node[above=of img, node distance=0cm, rotate=0, anchor=center,yshift=-0.9cm,xshift=0cm] {\scriptsize{Number of new researchers into 2025}};
     \end{tikzpicture}
    \end{minipage} & 
    \begin{minipage}{0.45\textwidth}
    \centering
    \begin{tikzpicture}
      \node (img)  {\includegraphics[scale=0.38]{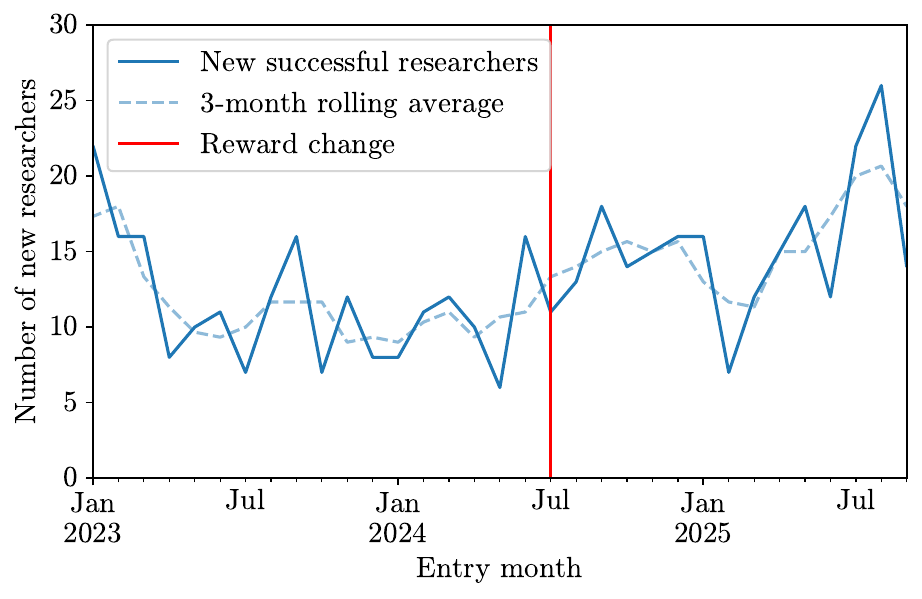}};
      \node[above=of img, node distance=0cm, rotate=0, anchor=center,yshift=-0.9cm,xshift=0cm] {\scriptsize{Number of new successful researchers into 2025}};
     \end{tikzpicture}
    \end{minipage} 
    \end{tabular}
    \caption{Number of new researchers through September, 2025. \textit{(Left)} Number of new researchers each month who submit a report for the first time to the treated program. \textit{(Right)} Number of new researchers each month who submit a report for the first time to the treated program, and successfully find a product bug in their first six months of participation.}
    \label{fig:new_2025}
\end{figure}

Figure \ref{fig:new_2025} shows that the number of new researchers and new successful researchers grew in 2025. This could either be a delayed effect of the reward change, or could also be due to a lower barrier to entry for bug hunting as AI tools became more sophisticated and more widely adopted. 

% Figure \ref{fig:all_timeseries_2022-2024} shows all bugs per month received from 2022-2024. We analyze data from 2023 onward as this is when detailed bug type labels became available. We also expect lessened pandemic effects from 2022 onward.

% \subsubsection{Top researchers} Figure \ref{fig:top_bottom_researchers_nocloud_2024} shows that the productivity of top and bottom researchers both stay relatively similar when Cloud bugs are removed. Plausible explanations for this include that the announcement of the Cloud program appealed especially to top researchers, or that the large GAVRP grant drew top researcher attention away from regular non-Cloud bug finding in December 2024.

% This does not definitively show that the announcement of the Cloud program was the cause in the growth in top Cloud researcher productivity independently of the reward change, but we also cannot rule out the fact that the announcement could have had a higher impact on top researchers. Given that the rewards offered by the CVRP were the same amounts as the rewards offered by the GAVRP, it is possible that the announcement of the program corresponded with increased advertisement of the higher rewards, or fed into a preference from top researchers for a more specialized program.

\section{Comparative case study with permutation placebos}\label{app:comparative_case_study}

To complement the RDD analysis, we leverage data from the two untreated programs to compute the gaps in outcomes over time between the treated program and the untreated programs in a difference-in-differences (DiD) model. Given that there are only three programs, we do not employ large-sample methods, but instead follow a comparative case study analysis with permutation placebos, in the style of \citet{abadie2010synthetic}.
%While this does not produce robust estimators in the same way as the previous RDD designs, the comparison between the treated and untreated programs is still informative.}

Specifically, let $Y_{g}^t(0)$ denote program $g$'s potential outcome in period $t$ if untreated in all periods. Let $Y_{g}^{t}(1)$ denote program $g$'s potential outcome in period $t$ if untreated in periods before $t^{*}$, and treated in periods $t^{*}$ to $t$. We assume the reward increase has no effect on the bug counts beforehand, so $Y_{g}^{t}(1) = Y_{g}^t(0)$ if $t < t^{*}$. Let $D_{g}^t = 1$ indicate that the treatment is active for program $g$ in period $t$, and $D_{g}^t = 0$ otherwise. The goal is to estimate the effect of the reward increase on the treated program at time $t$,
\[\tau^{\text{DiD}}_t = \E[Y_{g}^{t}(1) - Y_{g}^{t}(0) | D_g^{t'} = 1 \text{ for all } t' \geq t^{*}]\]
for $g = \text{treated}$, and for $t \geq t^{*}$.
To do this, we fit a dynamic two-way fixed-effects model:
\[Y_{g}^t = \alpha_{g} + \phi_{t} + D_{g}^t \beta^t + \varepsilon_{g}^t, \]
where $\alpha_{g}, \phi_{t}$ are parameters per group $g$ and period $t$, respectively, and $\varepsilon_{g}^t$ is a noise parameter. Estimates for $\hat{\tau}^{\text{DiD}}_t$ are given by the regression estimate for the coefficients $\beta^t$.

As permutation placebos, we also provide estimates for $\beta^t$ when the treatment assignment is swapped between the treated program and each of the two untreated programs. We intend for these to serve as additional sources of visual validation, rather than formal statistical inference. 
%each of the untreated programs is used as a stand-in for the ``treated program,'' and the actual treated program is used as a stand-in for that ``untreated program.'' 
% The idea is that these permuted estimates should not also show the same effects. 
%though formal permutation tests have been done with more untreated units \citep{mackinnon2020randomization}.

\subsection{Results with permutation placebos}

Figure \ref{fig:all_timeseries_app} \textit{(left)} shows all bugs per month received, from the treated and untreated programs, including permutation placebos.
% $\hat{\tau}^{\text{DiD}}$ takes a similar value, indicating that perhaps much of this change can be attributable to the reward change in the treated period.
% While the mean change in bugs per month and the Chow test statistics had p-value less than 0.05, the confidence intervals for the robust RDD and RKD tests were quite wide. 
% The lack of significance in $\hat{\tau}^{\text{RDD}}$ could be attributable to a delay in the increase in received bugs after the reward change: bugs tend to take weeks to months to find, and even if a researcher were to increase their efforts in July 2024, the fruits of those efforts may not show up until later. 
Figure \ref{fig:all_timeseries_app} \textit{(middle)} shows that OLS regression fits before and after the reward change indicate a positive change in slope after the reward change; however, the robust RDD and RKD designs could not verify significance.
%statistical significance remains questionable given the robust RKD design, where the effect disappears with a smaller bandwidth.
When comparing the treated program to the untreated programs, Figure \ref{fig:all_timeseries_app} \textit{(right)} shows that the gap between the counts in the treated program and the untreated programs seems to increase after the reward change, and the same effect was not observed in the permutation placebos. Figure \ref{fig:highvalue_combined_permutation} shows permutation placebos for each high-value bug type.

\begingroup
\setlength{\tabcolsep}{-2pt} 
\begin{figure*}[!ht]
    \centering
    \begin{tabular}{ccc}
    \begin{minipage}{0.34\textwidth}
    \centering
    \begin{tikzpicture}
      \node (img)  {\includegraphics[scale=0.34]{figures/all_timeseries_combined.pdf}};
      \node[above=of img, node distance=0cm, rotate=0, anchor=center,yshift=-1cm,xshift=0.2cm] {\tiny{Bug counts in treated vs. untreated programs}};
     \end{tikzpicture}
    \end{minipage} & \begin{minipage}{0.34\textwidth}
    \centering
    \begin{tikzpicture}
      \node (img)  {\includegraphics[clip, trim={0 5pt 0 7.1pt}, scale=0.34]{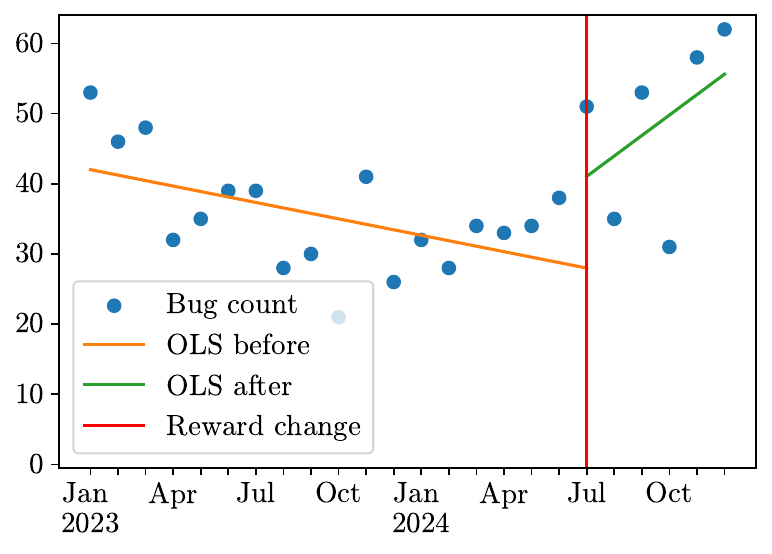}};
      \node[above=of img, node distance=0cm, rotate=0, anchor=center,yshift=-0.91cm,xshift=0cm] {\tiny{Regressions for treated program}};
    \end{tikzpicture}
    \end{minipage}& \begin{minipage}{0.34\textwidth}
    \centering
    \begin{tikzpicture}
      \node (img)  {\includegraphics[scale=0.34]{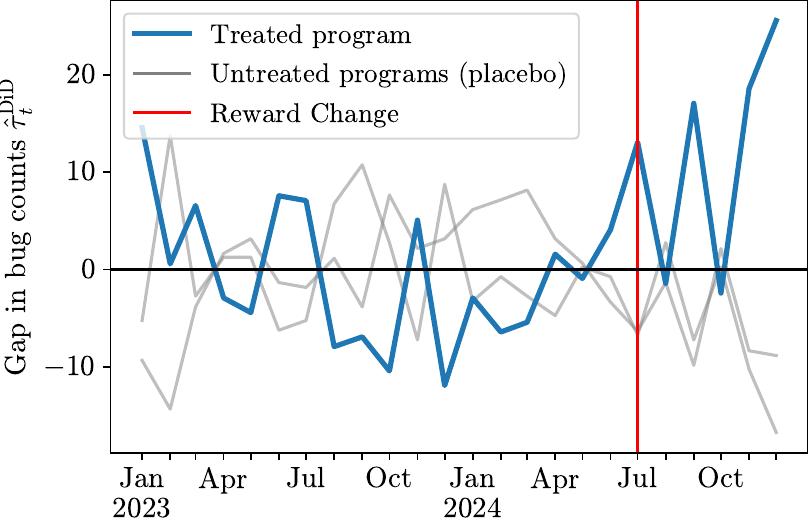}};
      \node[above=of img, node distance=0.3cm, rotate=0, anchor=center,yshift=-0.91cm,xshift=0.2cm] {\tiny{Gaps with permutation placebos}};
    \end{tikzpicture}
    \end{minipage}
    \end{tabular}
    \caption{
    % Visualizations of bug counts and regressions for all bugs in the treated and untreated programs. 
    \textit{(Left)} All bugs per month received in the treated and untreated programs. \textit{(Middle)} OLS regression fits in the treated program before and after the reward increase. \textit{(Right)} Gaps in bug counts between the treated program and untreated programs, normalized by the mean gap before treatment. The grey lines show permutation placebos for each untreated program.}
    \label{fig:all_timeseries_app}
\end{figure*}
\endgroup

\begingroup
\setlength{\tabcolsep}{-2pt} 
\begin{figure*}[!ht]
    \centering
    \begin{tabular}{ccc}
    \scriptsize{\underline{Bug counts in treated vs. untreated programs}} & \scriptsize{\underline{Regressions for treated program}} & \scriptsize{\underline{Gaps with permutation placebos}} \\
    \begin{minipage}{0.34\textwidth}
    \centering
    \begin{tikzpicture}
      \node (img)  {\includegraphics[scale=0.34]{figures/timeseries_tier0.pdf}};
      \node[above=of img, node distance=0cm, rotate=0, anchor=center,yshift=-1cm,xshift=0.2cm] {\scriptsize{Tier 0}};
     \end{tikzpicture}
    \end{minipage} & \begin{minipage}{0.34\textwidth}
    \centering
    \begin{tikzpicture}
      \node (img)  {\includegraphics[clip, trim={0 5pt 0 7.1pt}, scale=0.34]{figures/tier0_regressions_gc.pdf}};
      \node[above=of img, node distance=0cm, rotate=0, anchor=center,yshift=-0.91cm,xshift=0cm] {\scriptsize{Tier 0}};
    \end{tikzpicture}
    \end{minipage}& \begin{minipage}{0.34\textwidth}
    \centering
    \begin{tikzpicture}
      \node (img)  {\includegraphics[scale=0.34]{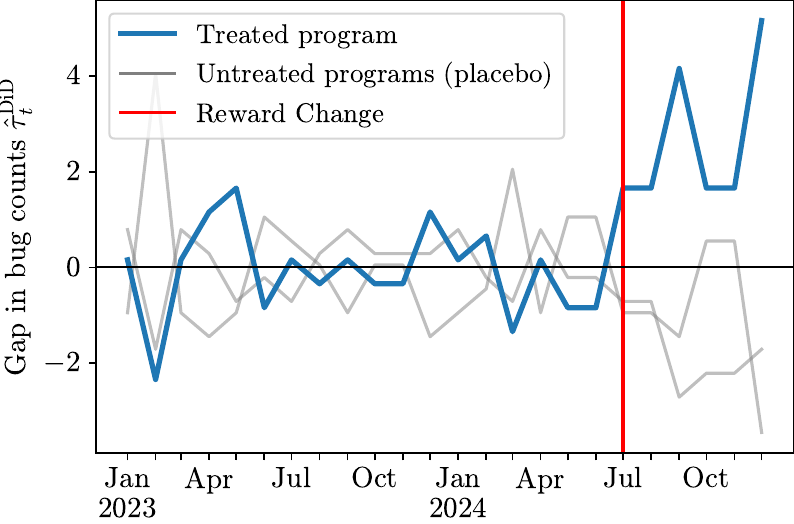}};
      \node[above=of img, node distance=0.3cm, rotate=0, anchor=center,yshift=-0.91cm,xshift=0.2cm] {\scriptsize{Tier 0}};
    \end{tikzpicture}
    \end{minipage} \\
    \begin{minipage}{0.34\textwidth}
    \centering
    \begin{tikzpicture}
      \node (img)  {\includegraphics[scale=0.34]{figures/timeseries_severityhigh.pdf}};
      \node[above=of img, node distance=0cm, rotate=0, anchor=center,yshift=-0.91cm,xshift=0.2cm] {\scriptsize{Severity $\geq$ High}};
     \end{tikzpicture}
    \end{minipage} & \begin{minipage}{0.34\textwidth}
    \centering
    \begin{tikzpicture}
      \node (img)  {\includegraphics[clip, trim={0 5pt 0 7.1pt}, scale=0.34]{figures/severity_high_regressions_gc.pdf}};
      \node[above=of img, node distance=0cm, rotate=0, anchor=center,yshift=-0.91cm,xshift=0cm] {\scriptsize{Severity $\geq$ High}};
    \end{tikzpicture}
    \end{minipage}& \begin{minipage}{0.34\textwidth}
    \centering
    \begin{tikzpicture}
      \node (img)  {\includegraphics[scale=0.34]{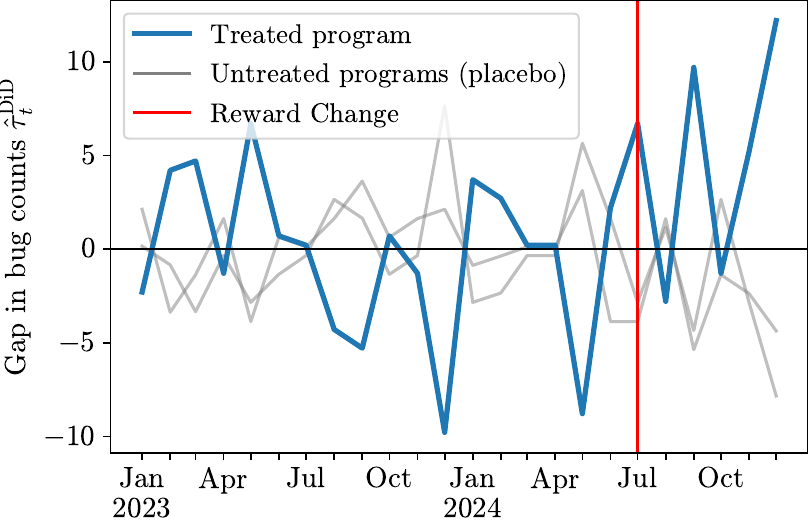}};
      \node[above=of img, node distance=0.3cm, rotate=0, anchor=center,yshift=-0.91cm,xshift=0.2cm] {\scriptsize{Severity $\geq$ High}};
    \end{tikzpicture}
    \end{minipage}\\
    \begin{minipage}{0.34\textwidth}
    \centering
    \begin{tikzpicture}
      \node (img)  {\includegraphics[scale=0.34]{figures/timeseries_merithigh.pdf}};
      \node[above=of img, node distance=0cm, rotate=0, anchor=center,yshift=-1cm,xshift=0.2cm] {\scriptsize{High Merit}};
     \end{tikzpicture}
    \end{minipage} & \begin{minipage}{0.34\textwidth}
    \centering
    \begin{tikzpicture}
      \node (img)  {\includegraphics[clip, trim={0 5pt 0 7.1pt}, scale=0.34]{figures/high_merit_rdd.pdf}};
      \node[above=of img, node distance=0cm, rotate=0, anchor=center,yshift=-0.91cm,xshift=0cm] {\scriptsize{High Merit}};
    \end{tikzpicture}
    \end{minipage}& \begin{minipage}{0.34\textwidth}
    \centering
    \begin{tikzpicture}
      \node (img)  {\includegraphics[scale=0.34]{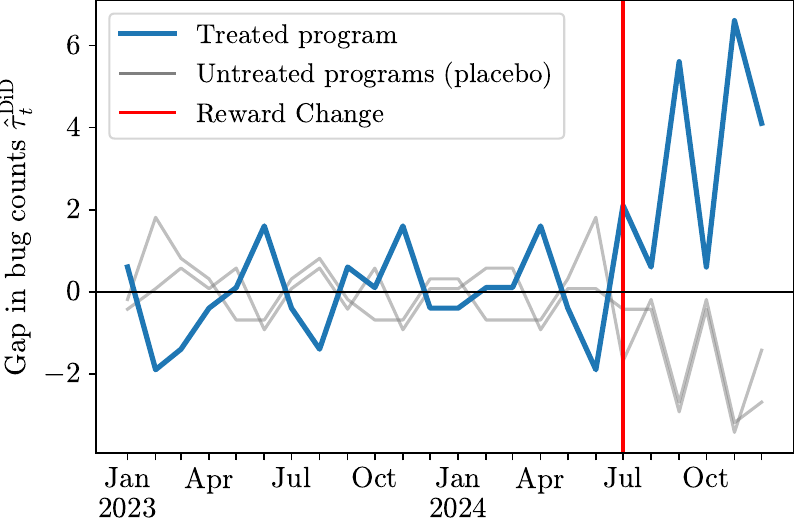}};
      \node[above=of img, node distance=0.3cm, rotate=0, anchor=center,yshift=-0.91cm,xshift=0.2cm] {\scriptsize{High Merit}};
    \end{tikzpicture}
    \end{minipage}
    \end{tabular}
    \caption{
    % Visualizations of bug counts and regressions for all bugs in the treated and untreated programs. 
      \textit{(Left column)} Bugs per month received in the treated and untreated programs for each of the high-value types. \textit{(Middle column)} OLS regression fits in the treated program before and after the reward increase for each of the high-value types. \textit{(Right column)} Gaps in bug counts between the treated program and untreated programs, normalized by the mean gap before treatment, for each of the high-value types.}
    \label{fig:highvalue_combined_permutation}
\end{figure*}
\endgroup

\end{document}